\documentclass[a4paper]{article}
\pdfoutput=1
\usepackage{authblk}
\usepackage{orcidlink}
\usepackage{graphicx} % Required for inserting images
\usepackage{pgffor}
\usepackage{geometry}
\usepackage{booktabs}
\usepackage{amsmath}
\usepackage{xcolor}
\usepackage{hyperref}
\usepackage[sorting=none,style=numeric-comp,giveninits=true,maxnames=1,backend=biber]{biblatex}
\usepackage{xspace} 
\usepackage{upgreek}

\def\MagUp {\mbox{\em Mag\kern -0.05em Up}\xspace}

 \def\Pmu         {\ensuremath{\mu}\xspace}

 \def\Ppi         {\ensuremath{\pi}\xspace}

 \def\Ppsi        {\ensuremath{\psi}\xspace}                 
                  
 \mathchardef\PDelta="7101
 \mathchardef\PXi="7104
 \mathchardef\PLambda="7103
 \mathchardef\PSigma="7106
 \mathchardef\POmega="710A
 \mathchardef\PUpsilon="7107
 \mathchardef\PPi="7105
 \def\PB      {\ensuremath{B}\xspace}                 
 \def\PD      {\ensuremath{D}\xspace}                 
 \def\PJ      {\ensuremath{J}\xspace}                 
 \def\PK      {\ensuremath{K}\xspace}                 
 \def\Pb      {\ensuremath{b}\xspace}                 
 \def\Pc      {\ensuremath{c}\xspace}

 \def\Ps      {\ensuremath{s}\xspace}

 \def\thebaroffset{0.18em}
\newcommand{\offsetoverline}[2][\thebaroffset]{\kern #1\overline{\kern -#1 #2}}%

\makeatletter
\ifcase \@ptsize \relax% 10pt
  \newcommand{\miniscule}{\@setfontsize\miniscule{4}{5}}% \tiny: 5/6
\or% 11pt
  \newcommand{\miniscule}{\@setfontsize\miniscule{5}{6}}% \tiny: 6/7
\or% 12pt
  \newcommand{\miniscule}{\@setfontsize\miniscule{5}{6}}% \tiny: 6/7
\fi
\makeatother

\DeclareRobustCommand{\optbar}[1]{\shortstack{{\miniscule (\rule[.5ex]{1.25em}{.18mm})}
  \\ [-.7ex] $#1$}}

\def\mup        {{\ensuremath{\Pmu^+}}\xspace}
\def\mun        {{\ensuremath{\Pmu^-}}\xspace} % muon negative (\mum is taken)

\def\mumu       {{\ensuremath{\Pmu^+\Pmu^-}}\xspace}

\def\ellm       {{\ensuremath{\ell^-}}\xspace}
\def\ellp       {{\ensuremath{\ell^+}}\xspace}

\def\squark    {{\ensuremath{\Ps}}\xspace}

\def\cquark    {{\ensuremath{\Pc}}\xspace}
\def\cquarkbar {{\ensuremath{\overline \cquark}}\xspace}
\def\ccbar     {{\ensuremath{\cquark\cquarkbar}}\xspace}
\def\bquark    {{\ensuremath{\Pb}}\xspace}

\def\pion   {{\ensuremath{\Ppi}}\xspace}

\def\pim    {{\ensuremath{\pion^-}}\xspace}

\def\kaon    {{\ensuremath{\PK}}\xspace}
\def\KorKbar {\kern \thebaroffset\optbar{\kern -\thebaroffset \PK}{}\xspace}

\def\Kp      {{\ensuremath{\kaon^+}}\xspace}
\def\Km      {{\ensuremath{\kaon^-}}\xspace}

\def\Kstarz  {{\ensuremath{\kaon^{*0}}}\xspace}

\def\D       {{\ensuremath{\PD}}\xspace}

\def\DorDbar {\kern \thebaroffset\optbar{\kern -\thebaroffset \PD}\xspace}

\def\Dp      {{\ensuremath{\D^+}}\xspace}
\def\Dm      {{\ensuremath{\D^-}}\xspace}

\def\DpDm    {\ensuremath{\Dp {\kern -0.16em \Dm}}\xspace}

\def\B       {{\ensuremath{\PB}}\xspace}

\def\BorBbar {\kern \thebaroffset\optbar{\kern -\thebaroffset \PB}\xspace}
\def\Bz      {{\ensuremath{\B^0}}\xspace}

\def\Bd      {{\ensuremath{\B^0}}\xspace}

\def\BdorBdbar {\kern \thebaroffset\optbar{\kern -\thebaroffset \Bd}\xspace}

\def\Bs      {{\ensuremath{\B^0_\squark}}\xspace}

\def\BsorBsbar {\kern \thebaroffset\optbar{\kern -\thebaroffset \Bs}\xspace}

\def\jpsi     {{\ensuremath{{\PJ\mskip -3mu/\mskip -2mu\Ppsi}}}\xspace}
\def\psitwos  {{\ensuremath{\Ppsi{(2S)}}}\xspace}

\def\Y#1S{\ensuremath{\PUpsilon{(#1S)}}\xspace}

\def\Lz          {{\ensuremath{\PLambda}}\xspace}

\def\LorLbar     {\kern \thebaroffset\optbar{\kern -\thebaroffset \PLambda}\xspace}

\def\Lb           {{\ensuremath{\Lz^0_\bquark}}\xspace}

\newcommand{\decay}[2]{\mbox{\ensuremath{#1\!\to #2}}\xspace} 

\def\to                 {\ensuremath{\rightarrow}\xspace}

\def\qsq       {{\ensuremath{q^2}}\xspace}

\def\BdToKpimm    {\decay{\Bd}{\Kp\pim\mup\mun}}

\def\bsll     {\decay{\bquark}{\squark \ell^+ \ell^-}}

\def\AT#1     {\ensuremath{A_{\mathrm{T}}^{#1}}\xspace}           % 2

\def\C#1      {\ensuremath{\mathcal{C}_{#1}}\xspace}                       % 9
\def\Cp#1     {\ensuremath{\mathcal{C}_{#1}^{'}}\xspace}                    % 7
\def\Ceff#1   {\ensuremath{\mathcal{C}_{#1}^{\mathrm{(eff)}}}\xspace}        % 9  
\def\Cpeff#1  {\ensuremath{\mathcal{C}_{#1}^{'\mathrm{(eff)}}}\xspace}       % 7
\def\Ope#1    {\ensuremath{\mathcal{O}_{#1}}\xspace}                       % 2
\def\Opep#1   {\ensuremath{\mathcal{O}_{#1}^{'}}\xspace}                    % 7

\newcommand{\aunit}[1]{\ensuremath{\text{\,#1}}}       
\newcommand{\tev}{\aunit{Te\kern -0.1em V}\xspace}
\newcommand{\gev}{\aunit{Ge\kern -0.1em V}\xspace}
\newcommand{\mev}{\aunit{Me\kern -0.1em V}\xspace}
\newcommand{\kev}{\aunit{ke\kern -0.1em V}\xspace}
\newcommand{\ev}{\aunit{e\kern -0.1em V}\xspace}
 
\newcommand{\mevc}{\ensuremath{\aunit{Me\kern -0.1em V\!/}c}\xspace}
\newcommand{\gevc}{\ensuremath{\aunit{Ge\kern -0.1em V\!/}c}\xspace}
\newcommand{\mevcc}{\ensuremath{\aunit{Me\kern -0.1em V\!/}c^2}\xspace}
\newcommand{\gevcc}{\ensuremath{\aunit{Ge\kern -0.1em V\!/}c^2}\xspace}
\newcommand{\gevgevcccc}{\ensuremath{\gev^2\!/c^4}\xspace} % for q^2

\def\deriv {\ensuremath{\mathrm{d}}}

\def\gsim{{~\raise.15em\hbox{$>$}\kern-.85em
          \lower.35em\hbox{$\sim$}~}\xspace}
\def\lsim{{~\raise.15em\hbox{$<$}\kern-.85em
          \lower.35em\hbox{$\sim$}~}\xspace}

\def\sPlot{\mbox{\em sPlot}\xspace}

\def\tell1  {TELL1\xspace}
\def\ukl1   {UKL1\xspace}

\newcommand{\ie}{\mbox{\itshape i.e.}\xspace}

\newcommand{\lhcborcid}[1]{\href{https://orcid.org/#1}{\hspace*{0.1em}\raisebox{-0.45ex}{\includegraphics[width=1em]{figs/orcidIcon.pdf}}}}

\def\th{{\ensuremath{\theta_h}}\xspace}
\def\tl{{\ensuremath{\theta_\ell}}\xspace}
\def\mkpi{{\ensuremath{m(K\pi)}}\xspace}

\title{Non-parametric and continuous extraction of amplitudes in realistic electroweak penguin decays}
\author{Tingzhen Xiao\thanks{tinxiao@ethz.ch}~~\orcidlink{0009-0004-2498-1680}}
\author{Anja Beck\thanks{anbeck@mit.edu}~~\orcidlink{0000-0003-4872-1213}}
\author{Eluned Smith\thanks{eluned@mit.edu}~~\orcidlink{0000-0002-9740-0574}}
\affil{\it\small Department of Physics and Laboratory for Nuclear Science, MIT, Cambridge, 02139, MA, USA}
\date{\today}

\begin{document}

\maketitle

\begin{abstract}
A novel approach to extract decay amplitudes in $B\to V(\to M_1M_2)\ellp\ellm$ processes, where $V$ represents a meson with either $J = 0$ (S-wave) or $J = 1$ (P-wave), was recently proposed. In this method, the dependence of the amplitudes on the dihadron and dimuon invariant masses is extracted model-independently through a fit to the helicity angles and a subsequent calculation of \sPlot weights. In this paper, we show that this method still performs well when applied to data with experimental realism, namely in the presence of detection efficiencies and background candidates, both of which can have non-factorising dependencies between the fitted helicity angles and extracted two-particle masses. We show that a non-factorising efficiency can be accommodated with a small change to the conventional calculation of the \sPlot weights, and a non-factorising background by expanding the description of the background distribution into individually factorising terms.

We illustrate the method using simulated \BdToKpimm data, containing both S- and P-wave contributions to the $\Kp\pim$ system, mixed with non-factorising background candidates and generated with a realistic function for the detection efficiency. We also show that this method can be applied over a much larger dimuon invariant-mass range than studied previously. This work suggests that the method is suitable for application to real experimental data, where it could provide model-independent decay-amplitude shapes for direct comparison with theoretical predictions. A measurement using this technique can improve sensitivity to potential new-physics effects in \bsll transitions, as well as the understanding of hadronic form factors, particularly in the S-wave system.

\end{abstract}

\section{Introduction}
%%%b->sll
Electroweak penguin decays, and in particular \bsll transitions, provide great opportunities to search for new physics.
The most well-studied example of such transitions, $\Bz\to \Kstarz\mumu$, exhibits tensions between measurement and Standard Model (SM) predictions~\cite{CMS:2024atz,Belle:2016fev,ATLAS:2018gqc,LHCb-PAPER-2020-002} that have been confirmed through recent results using the full statistical power of the Run 1 and 2 data sets collected by the LHCb experiment~\cite{LHCb:2025mqb,LHCb:2026enw}.
Several other decays and observables of similar \bsll processes have been measured~\cite{LHCb-PAPER-2015-023,LHCb-PAPER-2021-014, LHCb-PAPER-2014-006,LHCb-PAPER-2020-041,LHCb-PAPER-2023-033,LHCb-PAPER-2024-011,CMS:2024syx,CMS:2024atz, BaBar:2012mrf, Belle:2019xld,CMS:2024syx,LHCb-PAPER-2014-024,LHCb-PAPER-2017-013,LHCb-PAPER-2019-040,LHCb-PAPER-2021-004,LHCb-PAPER-2021-038,LHCb-PAPER-2022-045,LHCb-PAPER-2022-046,LHCb:2026dbi,CMS-PAS-BPH-23-003,CMS-PAS-BPH-24-005} and often deviate in similar ways from their SM prediction.

%%%Interpretations
Global analyses including different \bsll channels and observables reveal sizeable differences in the Wilson coefficient $\mathcal{C}_9$ with respect to its value predicted by the SM~\cite{Altmannshofer:2014rta,Capdevila:2017bsm,Beaujean:2013soa,Descotes-Genon:2013wba,Alguero:2021anc,Alguero:2023jeh}.
Any connection between an experimental observation and the Wilson coefficients relies strongly on an accurate model ---including appropriate uncertainties--- of the hadronic physics accompanying the \bsll transition.
For example, it is possible that the deviations may be due to underestimated uncertainties associated with non-local charm-loop processes mediated via $\bquark\to\squark\left[\ccbar\to\gamma^\ast\to\ellp\ellm\right]$,~\cite{Khodjamirian:2010vf,Lyon:2014hpa,Descotes-Genon:2013wba,Ciuchini:2015qxb,Gubernari:2020eft,Gubernari:2022hxn}.
A definitive interpretation of the tensions is only possible once the hadronic effects are understood properly, which can be aided via a detailed analysis of the \qsq-spectrum.

%%%Binned measurements
Most commonly, information about the \qsq-spectrum is obtained by measuring \bsll observables integrated over several ranges in \qsq, see e.g. Refs.~\cite{LHCb-PAPER-2015-051,LHCb-PAPER-2016-012,LHCb-PAPER-2020-002,LHCb-PAPER-2020-041,LHCb:2025mqb}.
Such binned measurements provide a model-independent proxy for the \qsq-dependence of an observable but are limited by the finite bin size of typically 1--2\gevgevcccc.
%%%Model-dependent unbinned measurements
In order to understand the dependence and the different contributions in more detail, the LHCb collaboration performed measurements of \BdToKpimm decays parametrising the \qsq dependence of the decay amplitudes to allow direct access to the Wilson coefficients~\cite{LHCb-PAPER-2024-011,LHCb-PAPER-2023-033,LHCb-PAPER-2023-032,LHCb:2026enw}.
While these measurements offer unique data-driven constraints on the size of e.g. non-local contributions, they rely on specific models for the hadronic form factors to describe the decay amplitudes which complicates the reinterpretation of the measurement using different models.

%%%Dihadron states
What is more, the dihadron spectrum in $B\to V(\to M_1M_2)\ellp\ellm$ typically consists of a rather narrow vector-meson resonance (P-wave) on top of a broader structure in a scalar (S-wave) configuration.
Higher partial waves and presence of both parities with the same spin are possible.
While all states are equally interesting in principle, theoretical and experimental studies of \BdToKpimm typically focus on the $\Kstarz(892) (\to \Kp \pim )$ vector contribution, denoted \Kstarz in the following.
This state can only be isolated by modelling the dihadron spectrum invoking sizeable systematic uncertainties in measurements due to the unknown shape of the underlying contributions.

%%% sPlotttttingggg for the win
We showed in Ref.~\cite{Beck:2025qxx} that a non-conventional rearrangement of the two-dimensional angular decay rate given by the hadron and lepton helicity angles allows to determine the magnitude of the S-wave and the transverse and longitudinal amplitude contributions to the P-wave.
A subsequent application of the \sPlot technique~\cite{Pivk:2004ty} provides the dependence of each contribution on the dihadron invariant mass \mkpi and the dimuon invariant-mass squared \qsq in a truly model-independent and continuous fashion.
The \mkpi shape could lead to a significant reduction of systematic uncertainties in future measurements by providing the exact shape of the S-wave contribution.
The \qsq shape could be vital input to the understanding of the decay composition and the interpretation of future measurements.
In this proof-of-principle study, we focused on the idealised case of a perfect detector and no background.
However the shape of the efficiency is typically highly non-factorising between the lepton/hadron helicity angle and the lepton/hadron invariant-mass.
Similarly, specific sources of background can result in contributions with strong correlations between the four-body invariant-mass, the dimuon invariant-mass squared, and the lepton helicity angle.
This paper discusses how to modify the approach such that the amplitudes can be extracted even in a realistic setting.
In addition, the amplitudes are now extracted in a significantly wider \qsq range reaching up to 19\gevgevcccc.

%%%Stucture
The contents are structured as follows.
We begin with providing an overview of the method presented in Ref.~\cite{Beck:2025qxx}, followed by a theoretical discussion of the presence and treatment of non-uniform efficiency effects and background.
Then, we validate our approach on both a large toy sample and a toy sample of realistic size including a comparison to a measurement of the method of moments.
Afterwards, we briefly examine how such a result could be presented.
We conclude by discussing the possible applications of this method as well as limitations.

\section{Setup}
%\textcolor{red}{I wouldn't have the sub-section headings in the actual paper, they are just here for easier debugging.}
%\subsection{Recap: decay rate}
%%%Decay rate
Assuming only scalar and vector contributions, the angular decay rate of \BdToKpimm after integrating over the angle between the hadron- and lepton-side decay planes can be expressed as
\begin{align}\label{eq:decrate}
\begin{split}
    \frac{1}{\Gamma_\text{total}}\frac{\deriv\Gamma}{\deriv\cos\th\deriv\cos\tl}
    &=f_\beta(\th,\tl) \underbrace{\left[3b_0^\text{P}+b_0^\text{S}\right]}_{(n_\beta)'} +f_0^\text{S}(\th,\tl) \underbrace{\left[n_0^\text{S}+\frac{3}{8}b_1^\text{P}+\frac{1}{2}b_0^\text{S}\right]}_{(n_0^\text{S})'} \\
    &+f_0^\text{P}(\th,\tl) \underbrace{\left[n_0^\text{P}-\frac{1}{8}b_1^\text{P}+\frac{1}{2}b_0^\text{P}\right]}_{(n_0^\text{P})'} +f_1^\text{P}(\th,\tl) \underbrace{\left[n_1^\text{P}+\frac{1}{2}b_1^\text{P}-2b_0^\text{P}\right]}_{(n_1^\text{P})'} \\
    &+\cos\tl\left(c_{10}+c_{11}\cos\th+c_{12}\cos^2\th\right) \\
    &+\cos\th\left(c_{1+}\left(\cos ^2\tl+1\right)+c_{1-}\left(\cos ^2\tl-1\right)\right) \\
     &= \mathcal{P}_\text{sig.}(\cos\th,\cos\tl) \ , \\
\end{split}
\end{align}
with the angular functions
\begin{align}\label{eq:angularfunc}
\begin{split}
    f_1^\text{P}(\th,\tl) &= \frac{9}{32} \left(\cos ^2\tl+1\right) \sin^2\th \ , \\
    f_0^\text{P}(\th,\tl) &= \frac{9}{8} \sin^2\tl \cos ^2\th \ , \\
    f_0^\text{S}(\th,\tl) &= \frac{3}{8} \sin^2\tl \ , \\
    f_\beta(\th,\tl) &= \frac{3}{16} \left(\cos ^2\tl+1\right) \ .
\end{split}
\end{align}
These functions are normalized to one such that the integral over the angular decay rate results in the constraint
\begin{align}
    \int\int\mathcal{P}_\text{sig.}(\cos\th,\cos\tl)~\deriv\cos\th~\deriv\cos\tl = (n^\text{P}_0)'+(n^\text{P}_1)'+(n^\text{S}_0)' + (n_\beta)'
    = 1 \ .
    \label{eq:norm}
\end{align}
The terms $n$, $b$, and $a$ are bilinear combinations of the transversity amplitudes where the superscript indicates which partial waves are involved and the subscript indicates which amplitudes are present.
The $n$ terms represent the absolute value squared of individual amplitudes, \ie
\begin{align}\label{eq:nterms}
\begin{split}
n_t^\text{A} &=  \beta ^2 |A_t| ^2 \ , \quad A=S,P \\
n_s^\text{A} &=  \beta ^2 |A_s| ^2 \ , \quad A=S,P \\
n_0^\text{A} &=  \beta ^2 \left(| A_0^L| ^2 + | A_0^R| ^2\right) \ , \quad A=S,P \\
n_1^\text{P} &=  \beta ^2 \left(| P_\parallel^L| ^2+| P_\perp^L| ^2+| P_\parallel^R| ^2+| P_\perp^R| ^2\right) \ . \\
\end{split}
\end{align}
In this notation, the helicity amplitudes for the P-wave are called $P^{L,R}_{\parallel,\perp,0,t,s}$ and the helicity amplitudes for the S-wave are $S^{L,R}_{0,t,s}$.
The subscript denotes the helicity configuration of the dimuon current (transverse, $\parallel,\perp$, longitudinal, $0$, timelike, $t$, or scalar $s$) and the superscript indicates whether the amplitude stems from a left- ($L$) or right-handed ($R$) current.
For better readability, the dependencies on the dihadron and dimuon invariant masses are kept implicit and the lineshape in the dihadron invariant-mass of the S- and P-wave have been absorbed into the amplitudes.
The common factor $\beta^2$ in all $n$ terms corresponds to the square of the speed of a lepton in the dilepton rest frame.
The $b$ terms,
\begin{align}\label{eq:bterms}
\begin{split}
b_0^\text{S} &= \frac{1}{2} \left(\frac{1-\beta^2}{\beta^2}\left(a_{\text{LR0}}^\text{S}+n_t^\text{S}+n_0^\text{S}\right)+n_s^\text{S}\right) \ , \\
b_0^\text{P} &= \frac{1}{2} \left(\frac{1-\beta^2}{\beta^2}\left(a_{\text{LR0}}^\text{P}+n_t^\text{P}+n_0^\text{P}\right)+n_s^\text{P}\right) \ , \\
b_1^\text{P} &= \frac{1-\beta^2}{\beta^2}\left(a_{\text{LR1}}^\text{P}+n_1^\text{P}\right) \ , \\
\end{split}
\end{align}
depend on the coefficients $n$ containing the amplitudes squared and interference terms between left- and right-handed amplitudes of the same partial wave which are
\begin{align}\label{eq:aterms}
\begin{split}
    a_{LR0}^\text{A} &= 2 \beta ^2 \Re\left[A_0^L (A_0^R)^*\right] \ , \quad A=S,P \ , \\
    a_{LR1}^\text{P} &= 2 \beta ^2 \Re\left[P_\parallel^L (P_\parallel^R)^*+P_\perp^L (P_\perp^R)^*\right] \ . \\
\end{split}
\end{align}
Note that the $b$-terms vanish in the Standard Model, where $A_s=0$, and when assuming massless leptons, $\beta\to1$.\footnote{All $a$ and $n$ terms carry a factor $\beta^2$, see Eqs.~\eqref{eq:nterms} and \eqref{eq:aterms}, resulting in a total suppression by $1-\beta^2$.}
The coefficients $c$ only appear with asymmetry terms and are nuisance parameters for this method as they vanish when integrating over the angles.
Explicit expressions for the $c$ terms can be found in App.~\ref{app:dwave} which outlines the decay rate including D-wave contributions.

The reader should be aware that the expression for the decay rate given in Eq.~\eqref{eq:decrate} differs from the one used in Ref.~\cite{Beck:2025qxx} which establishes the proof-of-principle for the method examined later.
In the preceding publication, the angular function $f_\beta$ was defined as
\begin{align*}
    f_\beta(\th,\tl) &= \frac{3}{8}r_\beta\sin^2\th+\frac{3}{4}(1-r_\beta)\cos^2\th \ ,
\end{align*}
where $r_\beta$ depends on the amplitudes.
This allows to absorb all $b$ terms into $n_\beta$ and $r_\beta$ such that the other coefficients correspond only to the sum of absolute values squared of amplitudes without \textit{contamination} due to $b$-terms, $(n_{0,1}^\text{S,P})' \to n_{0,1}^\text{S,P}$.
As shown in Ref.~\cite{Beck:2025qxx}, this method works with only a small bias.
Because $f_\beta(\th,\tl)$ is technically two angular functions, a fit of the angular decay rate needs to determine five coefficients, $n_0^\text{S}$, $n_0^\text{P}$, $n_1^\text{P}$, $r_\beta$ and $n_\beta$, when the angular space has only four degrees of freedom.
While this approach can produce acceptable results in regimes where $n_\beta$ is small, the fit and the resulting weights might generally suffer from some instability producing biased results in particular at low \qsq.
As a consequence, this follow-up study chose to define $f_\beta(\th,\tl)$ such that the angular functions represent a basis of the angular space.
The resulting fits are significantly faster due to the increased stability.

Ref.~\cite{Beck:2025qxx} shows that the dependence of the four coefficients $(n)'$, appearing in Eq.~\eqref{eq:decrate}, on \mkpi and \qsq can be extracted by fitting the angular dependence (referred to as fit variables) of the signal distribution $\mathcal{P}_\text{sig.}$ to data and then applying the \sPlot technique to extract the \qsq and \mkpi dependence (referred to as control variables).
The following two subsections discuss how this approach can work in the presence of non-trivial efficiency effects and complex backgrounds.

\subsection{Accounting for efficiency effects}
%%%Efficiency
In every real experiment detection and reconstruction limitations distort the observed distribution.
All these effects can usually be summarized in an efficiency function $\varepsilon$ that depends on the angles as well as the two-body and even the four-body invariant masses.
In order to account for the efficiency in our study, we perform a fit of the true distribution $\mathcal{P}_\text{sig.}(\cos\th,\cos\tl)$ to data weighted by the inverse of the efficiency $\varepsilon_i^{-1}$ at each data point.
In this case, the \sPlot weights $s_i$, extracting the individual contributions, can be calculated as before and the shapes are recovered correctly if the data is weighted by the \sPlot weight divided by the efficiency, $w_i=\tfrac{s_i}{\varepsilon_i}$.

It is important to note here that historically, the commonly used libraries used to calculate \sPlot weights, e.g. \texttt{hepstats}\cite{hepstats} and \texttt{sweights}\cite{sweights}, did not perform the calculation correctly in the presence of efficiency functions which do not factorize between the fit and control variables.
When calculating the \sPlot weighting function, several integrals of the following form are calculated,\footnote{See Eq.~(13) and its approximations in Ref.~\cite{Dembinski:2021kim} or Eq.~(11) in Ref.~\cite{Pivk:2004ty}.}
\begin{align}
    W_{ab} = \int \frac{g_a(x)g_b(x)}{\sum_i g_i(x)} \ .
\end{align}
Here, the functions $g_i(x)$ are individual components ---the $f_i(\th,\tl)n_i$ terms in our case--- of the total distribution $\sum_i g_i(x)$.
The integrals $W_{ab}$ are elements of the $W$ matrix required to compute the \sPlot weights, see Refs.~\cite{Dembinski:2021kim,Pivk:2004ty}.
This integral is commonly calculated through numerical integration using the data set employed in the fit.
This produces correct results in the case of 1) efficiency-weighted fits with either uniform efficiency or an efficiency that factorizes between the fit variables, i.e. the angular variables, and control variables, i.e. the diparticle masses, as well as 2) unweighted fits.
For a weighted fit with non-factorizing efficiency-weights, the integral needs to be calculated by weighting the integration terms with the inverse of the efficiency, i.e. importance sampling.\footnote{The authors of this paper are making an effort to update these software tools. At the time of publication, the most recent version of some libraries already include the necessary changes.}
A simple but typical example for the efficiency shape of a rare decay, $H_b\to h_1^+h_2^-\mumu$, observed at the LHCb experiment is assumed and displayed in Fig.~\ref{fig:efficiency} in App.~\ref{fig:efficiency}.

\subsection{Incorporating backgrounds}
%%%Background
Adding the four-body invariant-mass, $m$, to the fit allows to separate backgrounds present in the fit because the signal peaks at the \Bd mass while the background typically populates a wide range.
The three-dimensional distribution can be written as
\begin{align}
    \mathcal{P}_\text{total}(m,\cos\th,\cos\tl) = \mathcal{P}_\text{bkg.}(m)\mathcal{P}_\text{bkg.}(\cos\th,\cos\tl) N_\text{bkg.} + \mathcal{P}_\text{sig.}(m)\mathcal{P}_\text{sig.}(\cos\th,\cos\tl) N_\text{sig.} \ ,
\end{align}
where $N_\text{sig.}$ and $N_\text{bkg.}$ are the number of signal and background candidates, $\mathcal{P}_i(m)$ are the one-dimensional mass distributions, and $\mathcal{P}_i(\cos\th,\cos\tl)$ are the two-dimensional angular distributions.
The background sample used in this paper has a correlation of 17\% between the four-body invariant-mass and the dilepton invariant-mass.
All other background variables factorize.
This is a choice made for simplicity but there is no reason that prevents the method from being applied to distributions that include correlations between these variables as long as the fit model, $\mathcal{P}_\text{total}$, is accurate.
Figure~\ref{fig:background} in App.~\ref{app:bkg} shows the background samples obtained as the sum of one sample that represents fully factorizing combinatorial background and another background sample that has a strong correlation between the four-body and dimuon invariant-mass.

If the background factorizes between the three fit variables in $\mathcal{P}_\text{total}(m,\cos\th,\cos\tl)$ and the two control variables, \mkpi and \qsq, the \sPlot method can be applied right away.
This is for example the case for truly random combinations of particles.
However, there can be other more complex sources of background including contributions from partially reconstructed tree-level decays which can appear predominantly in the lower mass-sideband of the \mbox{\BdToKpimm} peak and at low \qsq.
%Due to the relationship between the two muons, these backgrounds produce a distinct peak close to $+1$ (or $-1$ depending on the definition of the angle).
Because the four-body invariant-mass and the dimuon mass are highly correlated in this example, the standard \sPlot procedure fails.
This can be rectified by approximating the correlated part of the distribution by the sum of individually factorizing terms as suggested in Ref.~\cite{Dembinski:2021kim},
\begin{align}\label{eq:expansion}
    \mathcal{P}_\text{bkg.}(m,\qsq) = \mathcal{P}_\text{bkg.}(m)\sum_{i=0}^\infty p_\text{bkg.}^i(m) p_\text{bkg.}^i(\qsq) \ .
\end{align}
Note that the dependence on \qsq is not known but extracted using the \sPlot weights.
Explicitly, we choose to model the four-body invariant-mass part of the distribution by the polynomials
\begin{align}
    p_\text{bkg.}^i(m) = (i+1)\left(\frac{m-m_\text{min}}{m_\text{max}-m_\text{min}}\right)^i \ ,
\end{align}
representing a monomial basis for the four-body invariant-mass scaled to the range $[0,1]$.
The factor $(i+1)$ normalizes the monomials as $\int_0^1x^i\deriv x=(i+1)^{-1}$.
It is straight-forward to compute \sPlot weights to extract each $p_\text{bkg.}^i(\qsq)$ term individually.
In the limit of $i\to\infty$, the sum of these individual weights recovers the total background contribution.
In practice however, the inclusion of more than a few polynomials results in numerical instabilities when inverting the $W$ matrix in the \sPlot weight computation due to increasingly smaller and correlated components.
For the study at hand, we use $i\leq4$.
Note that in our example, the probability distribution function describing the angular dependence is unaffected as there is no correlation between the angles and the control variables.

\section{Studies using pseudo-data}
%%%Model
In order to illustrate and test our method, toy data of \BdToKpimm decays is generated using local form factors from Ref.~\cite{Bharucha:2015bzk} and non-local effects parametrised as in Ref.~\cite{Altmannshofer:2014rta}.
The setup allows for massive leptons. The Wilson coefficients are set to a SM prediction obtained using the \texttt{flavio} software~\cite{Straub:2018kue}.
The S-wave lineshape is described using a LASS model~\cite{Lu:2011jm}.
The background toy is uniformly distributed in the two-body invariant-masses and has non-uniform angular dependence.
There is a highly non-factorizing term connecting the four-body invariant-mass and the dimuon invariant-mass squared, as previously discussed.

Both the signal and background toy are limited to the ranges $0.746<\mkpi<1.5\gevcc$ and $1.1<\qsq<19\gevgevcccc$.
Efficiency effects are introduced by resampling the signal and background using the accept-reject method with the efficiency function introducing additional non-factorizing dependencies between the lepton helicity angle and the dimuon invariant-mass.
In all following studies, two thirds of the data set correspond to signal and one third corresponds to background.

\begin{figure}
    \centering
    \includegraphics[width=.49\textwidth]{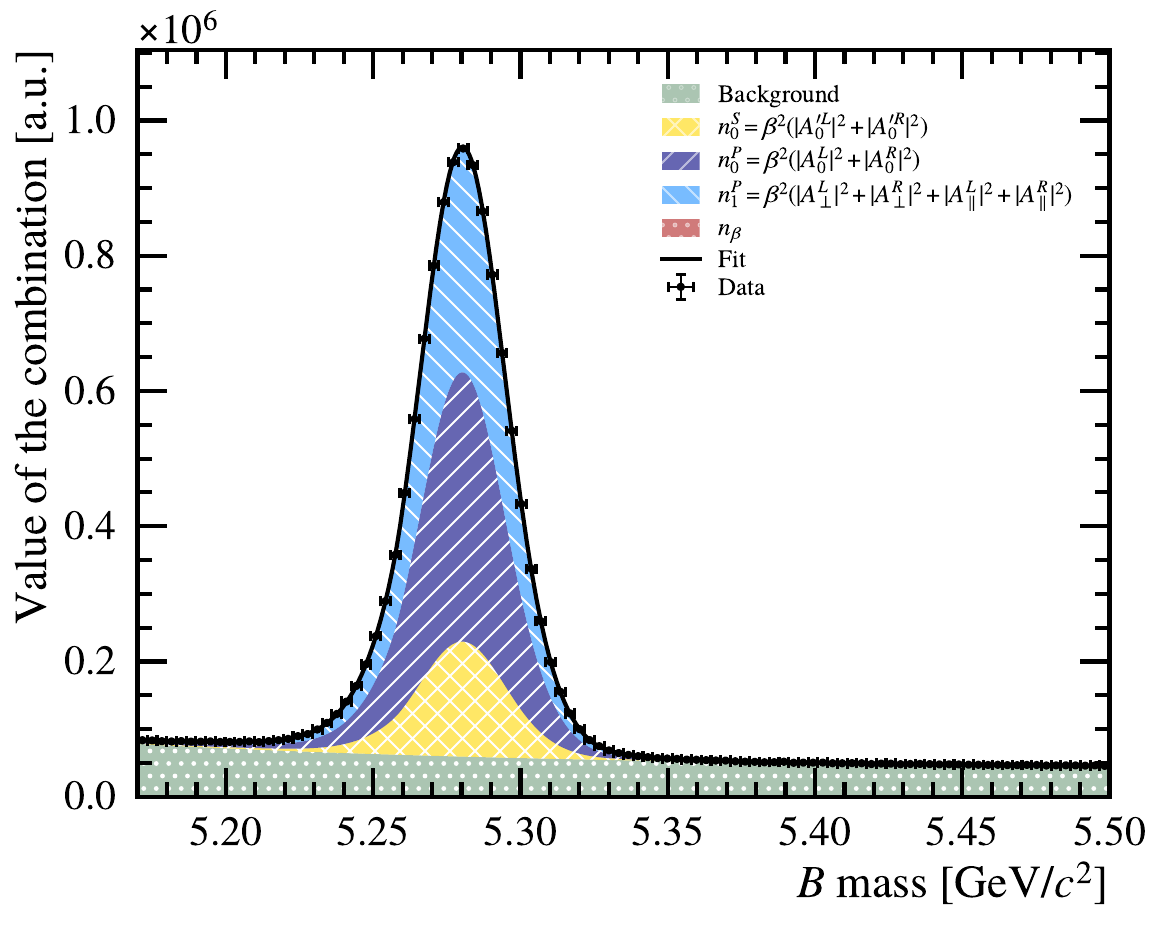}
    \includegraphics[width=.49\textwidth]{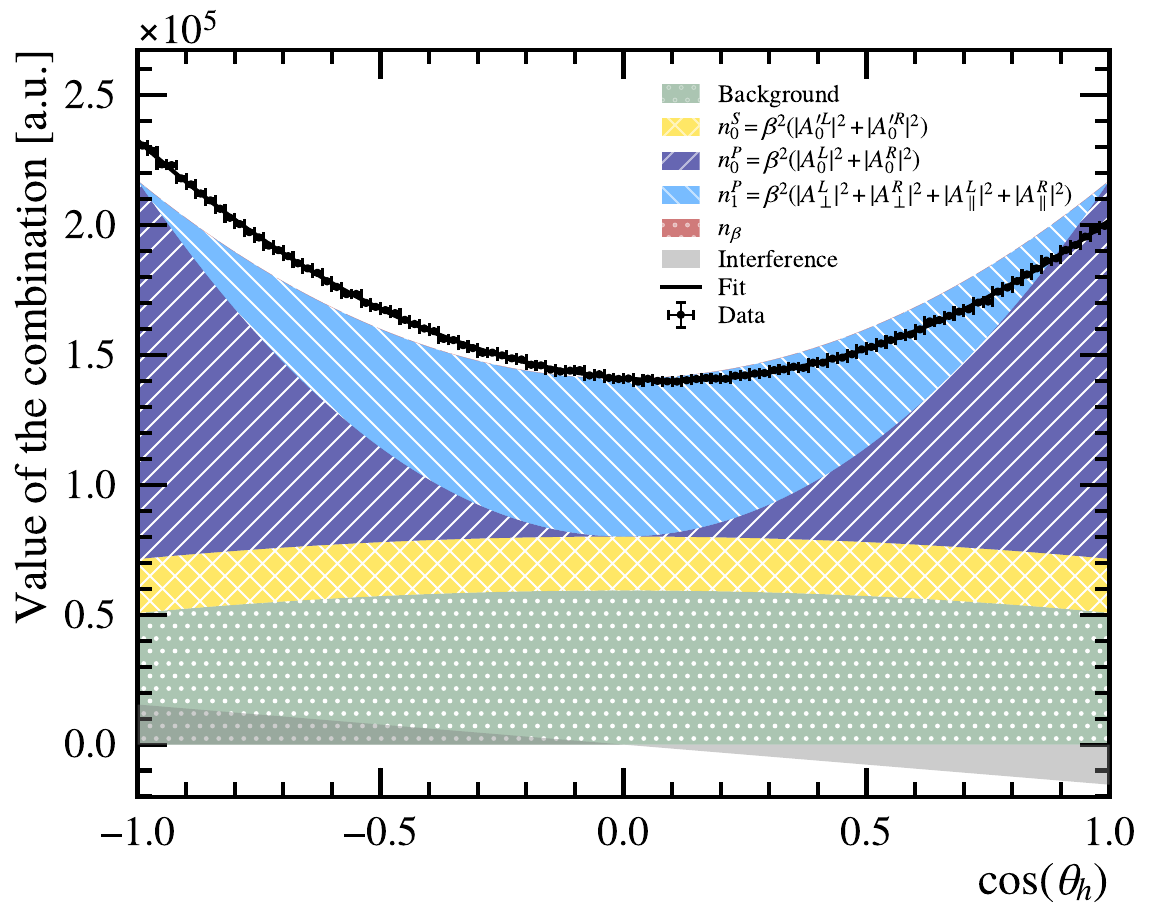}%
    \includegraphics[width=.49\textwidth]{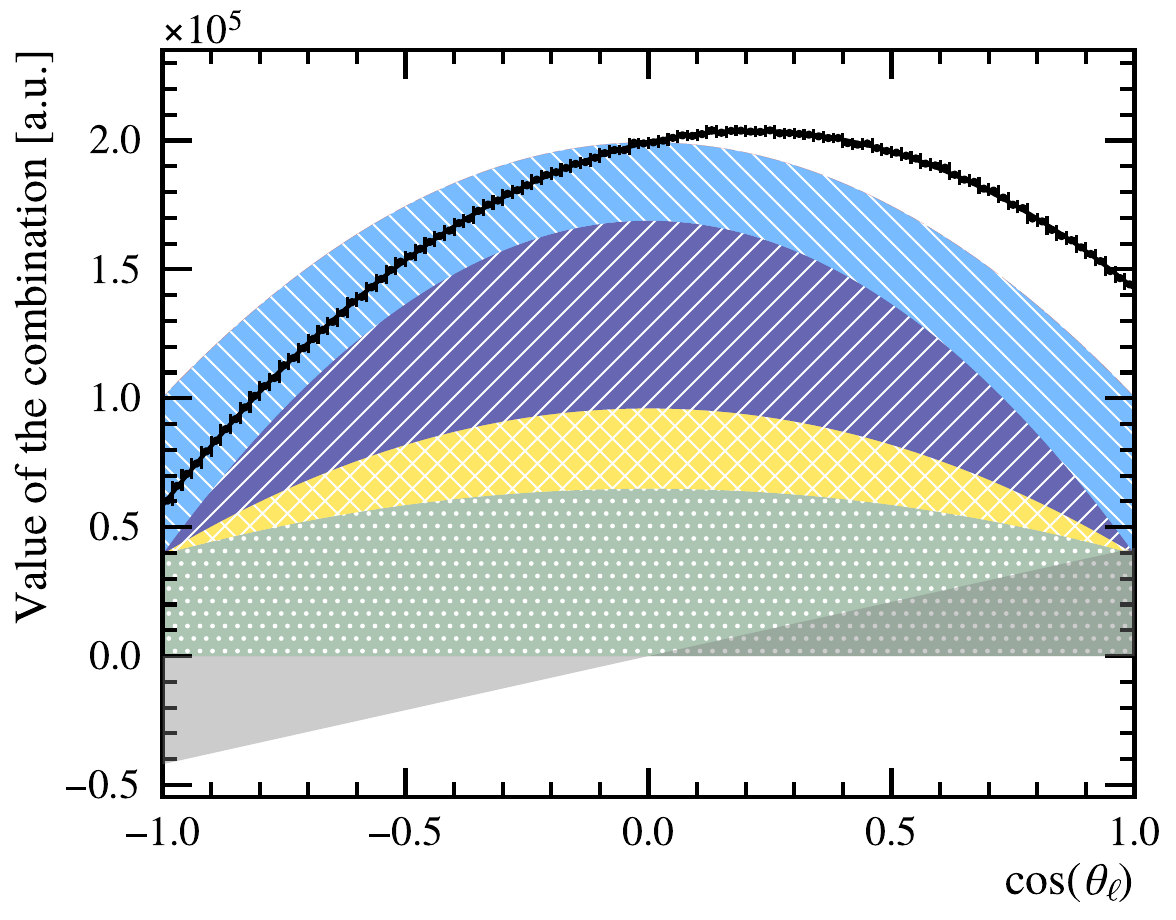}
    \caption{Projection of a large toy data set of more than 10 million data points, generated assuming massless muons, onto the (top) four-body invariant-mass, (left) hadron helicity angle and (right) muon helicity angle, weighted by the inverse of the efficiency. The fitted angular decay rate including all its components is also shown.}
    \label{fig:fitprojections_massless}
\end{figure}

%%%Large toys
Figure~\ref{fig:fitprojections_massless} shows the fit to a large toy data set with 18 million generated decays, projected onto the two angles and the four-body invariant mass.
The samples used in this figure are generated assuming massless muons.
Appendix~\ref{app:massive_muons} contains the equivalent figures for toy data with massive muons.
Note that the interference in the hadron helicity angle (bottom left plot) originates from interference between S- and P-wave amplitudes while the interference in the muon helicity angle (bottom right plot) originates from interference between the transverse P-wave amplitudes.

\begin{figure}
    \centering
    \includegraphics[width=0.49\textwidth]{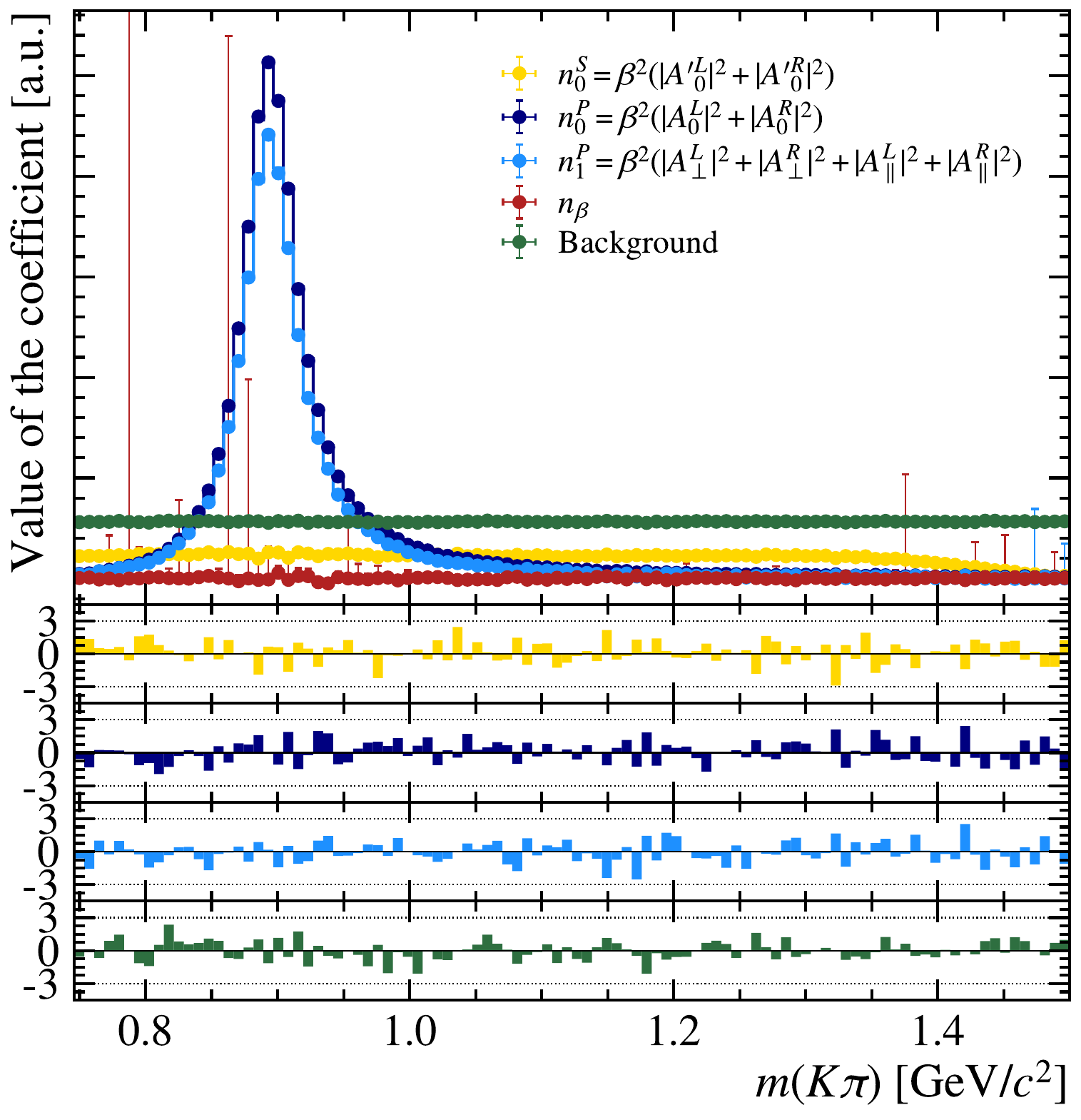}%
    \includegraphics[width=0.49\textwidth]{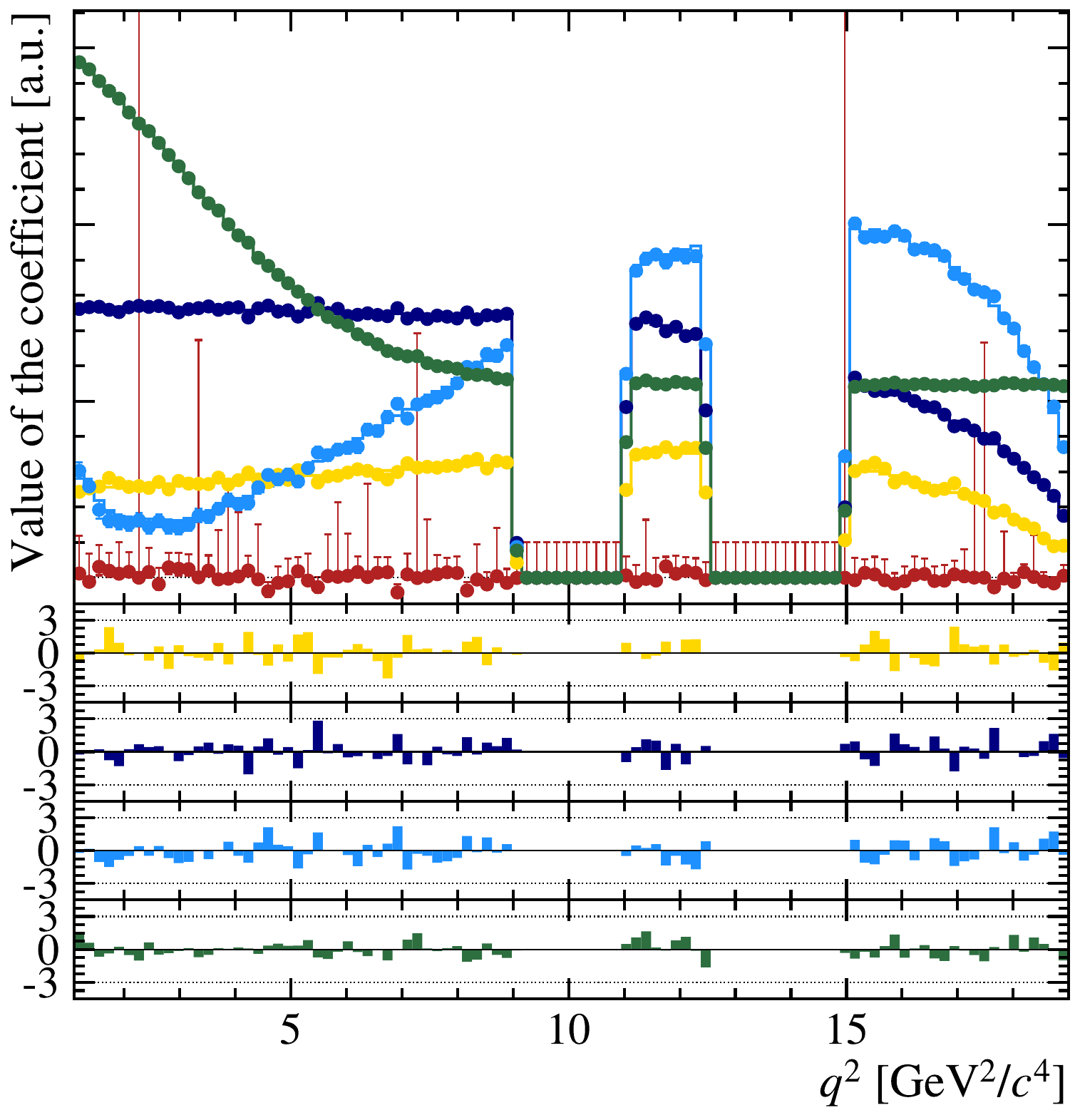}
    \caption{Extracted shapes (data points) compared to the generated shapes (lines) in the (left) dihadron invariant-mass and (right) dimuon invariant-mass squared for a large toy data set of more than 10 million data points, generated assuming massless muons.
    The three bottom plots show the bin-wise pulls for the three components in their respective colour.
    As $\beta\to1$ when assuming massless muons, the $(1-\beta^2)$ dependent term $n_\beta$ must be consistent with zero.}
    \label{fig:sweighted_highstats_massless}
\end{figure}

Calculating the \sPlot weights based on the combination of the four-body invariant mass distribution and the angular distribution allows to separate the individual components as shown in Fig.~\ref{fig:sweighted_highstats_massless}.
In addition to the weighted data set, this figure also shows reference samples plotted as lines.
As evidenced by direct comparison and the pulls in the bottom part of the plots, the extracted shapes (data points) are in agreement with the true distributions (lines) obtained by sampling from a model where only the amplitudes of interest are non-zero.
The regions around the \jpsi and \psitwos resonances have been vetoed before performing the fit and calculating the sWeights.
Note that the interference terms shown in gray in Fig.~\ref{fig:fitprojections_massless} vanish in the projections on the invariant-masses due to the asymmetry in the angular terms.

\begin{figure}
    \centering
    \includegraphics[width=0.49\textwidth]{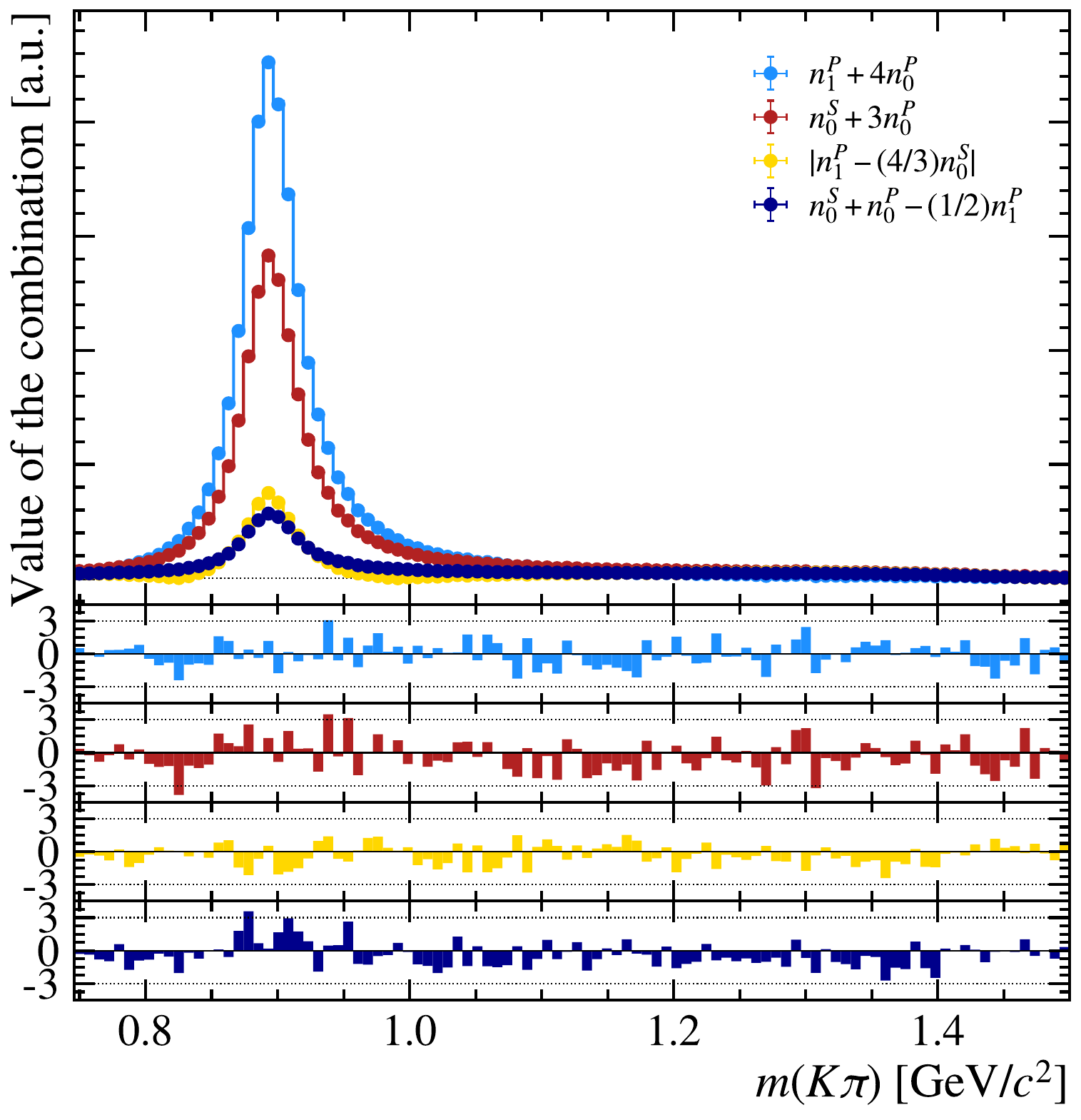}%
    \includegraphics[width=0.49\textwidth]{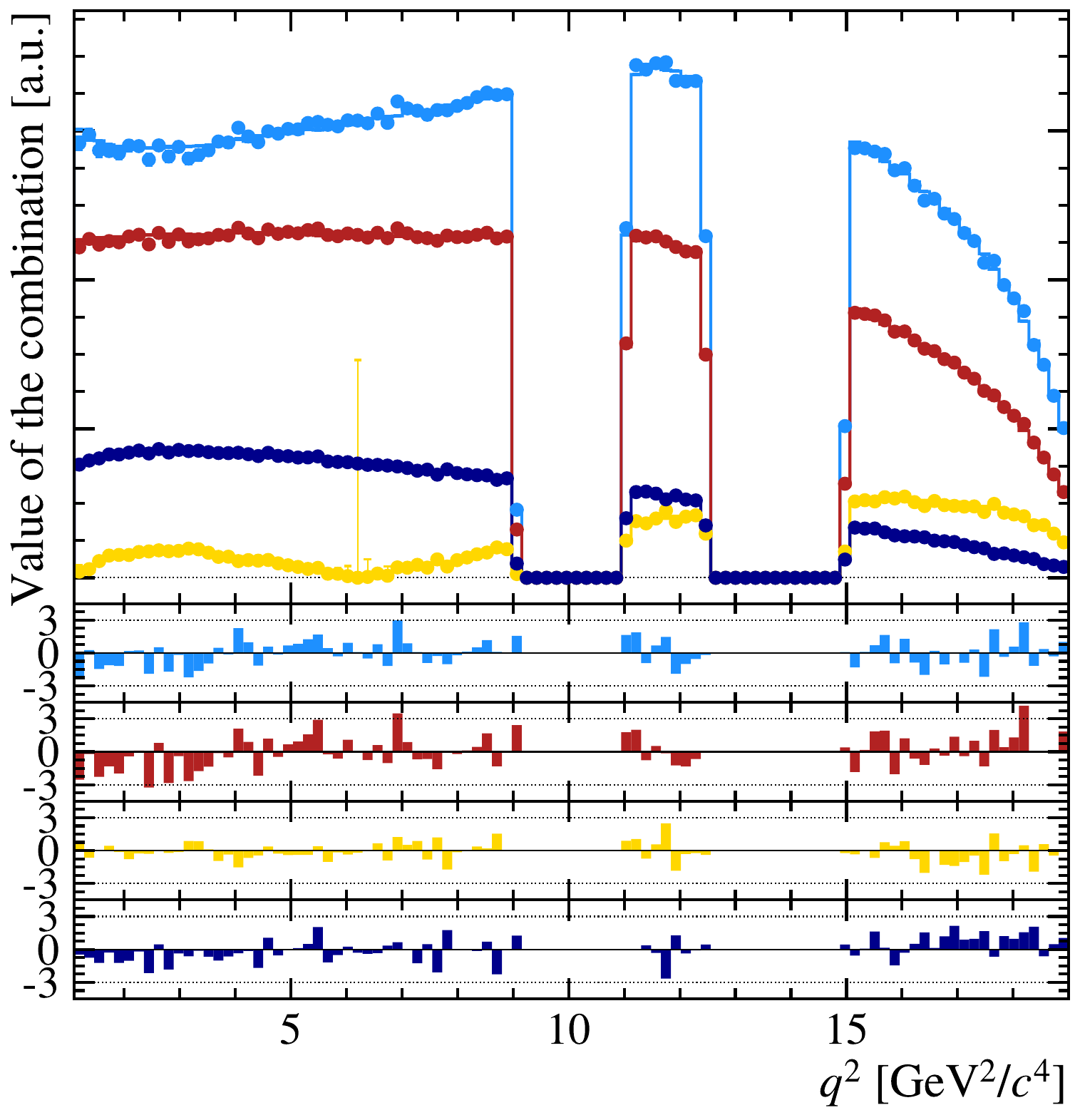}
    \caption{Shape combinations (data points) compared to the combined reference shapes (lines) in the (left) dihadron invariant-mass and (right) dimuon invariant-mass squared for a large toy data set of more than 10 million data points, generated assuming massive muons.
    The faint dotted horizontal line indicates zero.
    The four bottom plots show the bin-wise pulls for the four combinations in their respective colour.
    }
    \label{fig:alternative_highstats_massive}
\end{figure}

\begin{figure}
    \centering
    \includegraphics[width=0.49\textwidth]{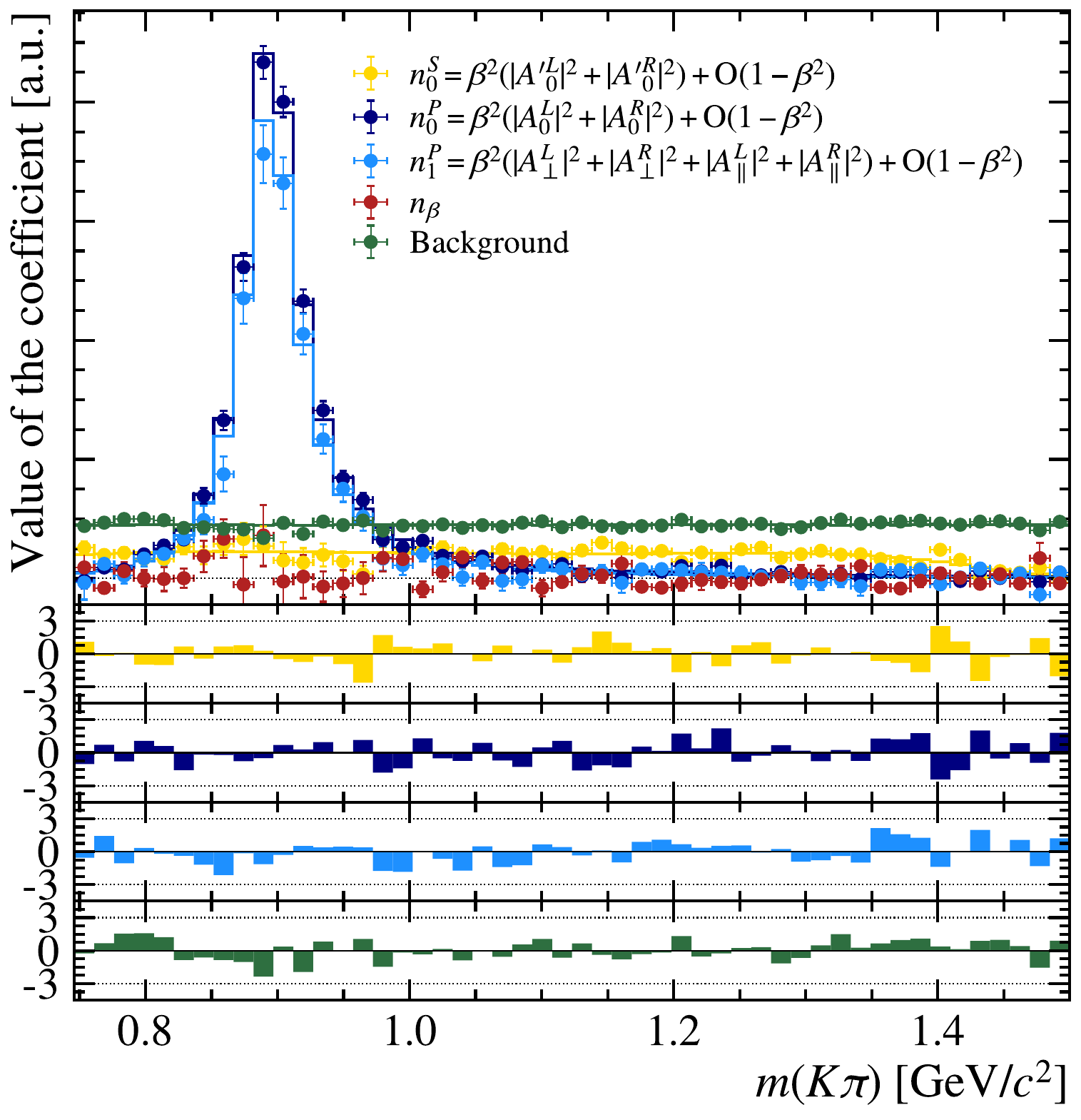}
    \includegraphics[width=0.49\textwidth]{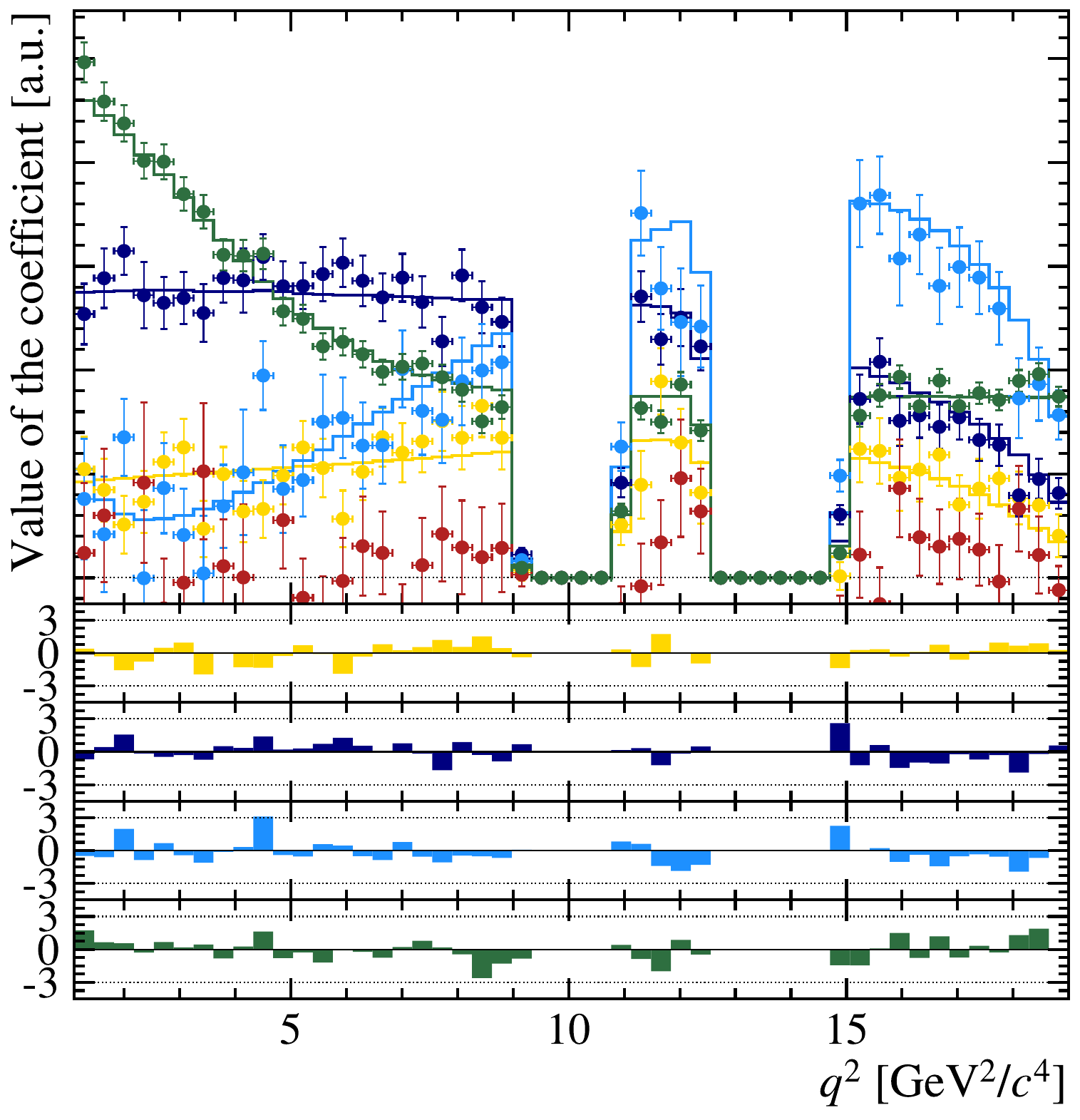}
    \caption{Statistical uncertainty on the shape of the four components in the (left) dihadron and (right) dimuon invariant-mass based on the chosen binning scheme in an example toy data set containing 90~000 data points.
    The faint dotted horizontal line indicates zero.
    No generated shape for $n_\beta$ is shown because the generation of a representative toy sample is non-trivial.}
    \label{fig:sweighted_nominal_massive}
\end{figure}

The reference samples are produced by generating toy data where all amplitudes are set to zero except $P_{\perp,\parallel}^{L,R}$ for $(n_1^P)'$, $P_0^{L,R}$ for $(n_0^P)'$, or $S_0^{L,R}$ for $(n_0^S)'$.
As a consequence, the reference samples cannot fully represent the additional $b$-terms appearing in the coefficients $(n)'$, see Eqs.~\eqref{eq:decrate} and \eqref{eq:bterms}.
In the case of massless muons considered so far, this is not relevant as the $b$-terms vanish.
Producing samples that truly represent the $(n)'$ coefficients for massive muons is non-trivial.
There are however combinations of coefficients where the $b$-terms cancel
\begin{align}\label{eq:alternative_checks}
\begin{split}
    &(n_1^P)'+4(n_0^P)'=n_1^P+4n_0^P \ , \quad
    (n_1^P)'-\frac{4}{3}(n_0^S)'=n_1^P-\frac{4}{3}n_0^S \ , \\
    &(n_0^S)'+3(n_0^P)'=n_0^S+3n_0^P \ , \quad
    (n_0^S)'+(n_0^P)'-\frac{1}{2}(n_1^P)'=n_0^S+n_0^P-\frac{1}{2}n_1^P \ .
\end{split}
\end{align}
These relationships allow to examine the shapes for biases even in the presence of massive muons when no perfect reference can be generated.\footnote{Note that the four relationships in Eq.~\eqref{eq:alternative_checks} are not linearly independent. It is not possible to rearrange them and obtain expressions for the $(n)'$ coefficients.}
Appendix~\ref{app:massive_muons} contains the equivalent of Fig.~\ref{fig:sweighted_highstats_massless} for massive muons illustrating the small biased introduced due to the incomplete $b$-terms in the reference samples.
We want to stress that this bias can be attributed entirely to the difficulty of generating reference samples that accurately represent the extracted coefficients $(n)'$ as opposed to a bias in the measured value.
We have high confidence in this statement because the bias disappears for massless muons and when considering the combinations in Eq.~\eqref{eq:alternative_checks} where the $b$-terms cancel.
What is more for the models employed in this study, the mismatch is only notable in data sets exceeding several million data points, far beyond a realistic data set collected by the LHCb experiment through the end of its run time.
A comparison between the sWeighted data set and the reference samples using the relationships in Eq.~\eqref{eq:alternative_checks} is shown in Fig.~\ref{fig:alternative_highstats_massive}.
Similarly to the massless example considered before, there are no localized biases confirming the reliability of the weights in a large sample.

%%%Small toys
Figure~\ref{fig:sweighted_nominal_massive} shows the extracted distributions obtained on a sample of realistic size, containing 90~000 \BdToKpimm candidates of which two thirds are signal and one third is background.
This corresponds roughly to the data collected by the LHCb collaboration to date after reasonable selections.
The per-bin uncertainties are calculated from the sum of weights squared.
The correct coverage of these uncertainties on the realistic-size toy is confirmed through a pull study of fits to 100 independent samples.
The deviation between the reference samples and the sWeighted data set in the massive muon case discussed before is invisible now due to the much larger uncertainties.

Note that all of the described studies have been performed using a few different models to generate signal and background decays which all produced consistent results.

\section{Comparison to a measurement of the angular moments}
%%%Traditional coefficients
In the massless lepton case, the extracted coefficients can be mapped to the more conventional $S_i$ angular observables~\cite{Altmannshofer:2008dz}, for example
\begin{align}\label{eq:angcoeffs}
    \left(\frac{\deriv\Gamma}{\deriv\qsq}\right)^{-1}n_0^\text{P} = -S_{2c} \quad\text{and}\quad \left(\frac{\deriv\Gamma}{\deriv\qsq}\right)^{-1}n_1^\text{P} = 4 S_{2s} \ .
\end{align}
When measuring angular coefficients using the method of moments~\cite{Beaujean:2015xea}, signal and background are disentangled through \sPlot weights $s_i$, calculated based on a fit to the four-body invariant-mass.
Typically, these fits are performed per \qsq bin which dilutes potential correlations.
In this study however, we perform the fit to the four-body invariant-mass on the full data and use the expansion trick to compute one set of signal sWeights for the entire range.
The signal-weighted and efficiency-corrected angular moment can then be calculated as the sum over all data points $i$ in a given \qsq bin
\begin{align}
\begin{split}
    S_{2s} &= \frac{1}{N}\sum_i \left[\frac{5}{6} \mathcal{L}_2(\cos\tl_i)\left(2\mathcal{L}_0(\cos\th_i) - 5\mathcal{L}_2(\cos\th_i)\right) \times \frac{s_i}{\varepsilon_i}\right] \ , \\
    S_{2c} &= \frac{1}{N}\sum_i \left[\frac{5}{3} \mathcal{L}_2(\cos\tl_i)\left(\mathcal{L}_0(\cos\th_i) + 5\mathcal{L}_2(\cos\th_i)\right) \times \frac{s_i}{\varepsilon_i} \right]\ ,
\end{split}
\end{align}
normalized by the effective number of candidates $N=\sum_i\frac{s_i}{\varepsilon_i}$.
The angular dependence is given in terms of Legendre polynomials $\mathcal{L}$.
The \sPlot weights $s_i$ are computed from an unweighted fit to the four-body invariant mass and $\varepsilon_i$ is the efficiency for a given data point.
Because the method of moments cannot easily distinguish between S- and P-wave, the following comparison between the \sPlot method and the method of moments is performed using a signal sample without S-wave contributions.

\begin{figure}
    \centering
    \includegraphics[width=.49\textwidth]{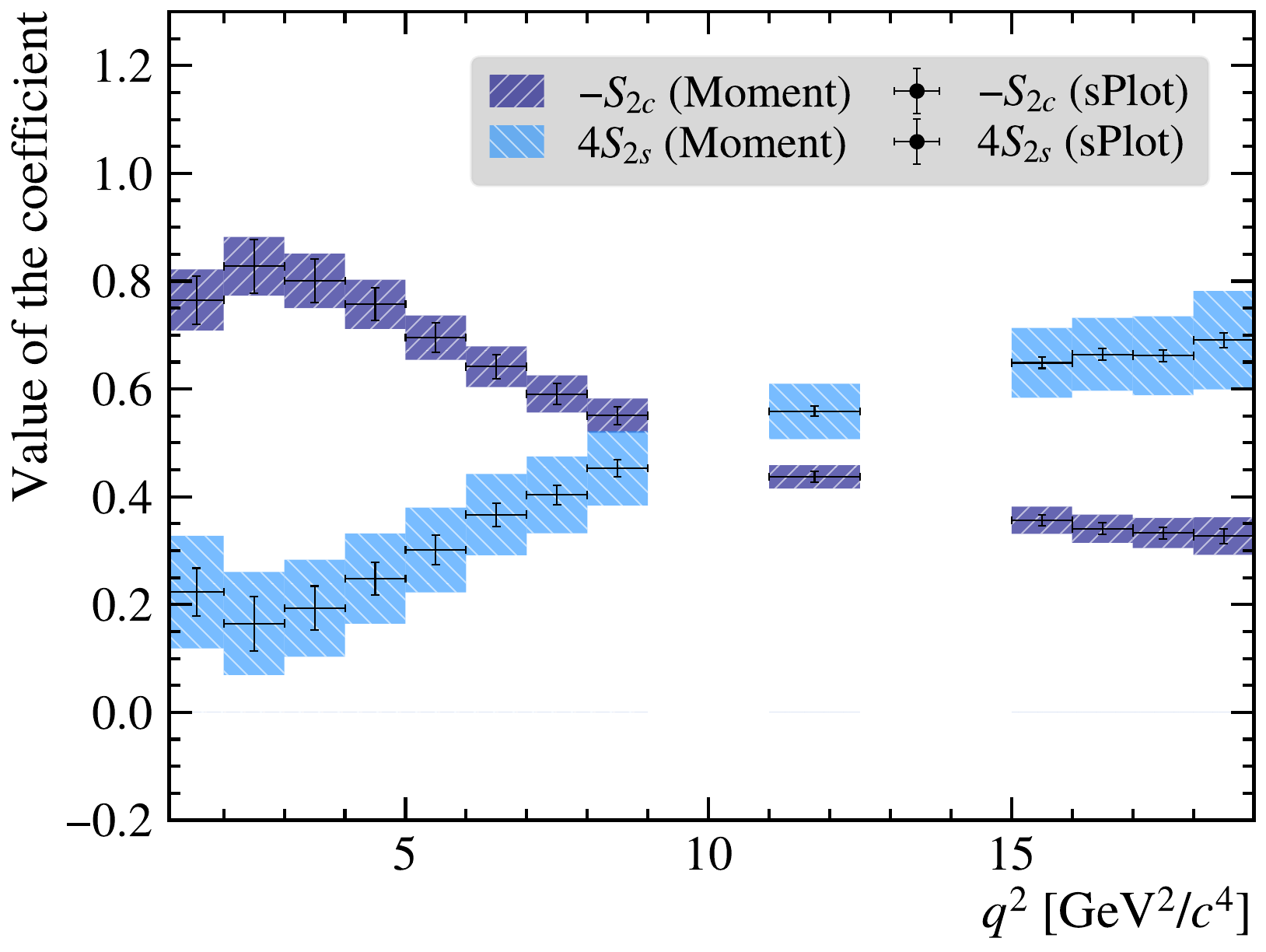}
    \includegraphics[width=.49\textwidth]{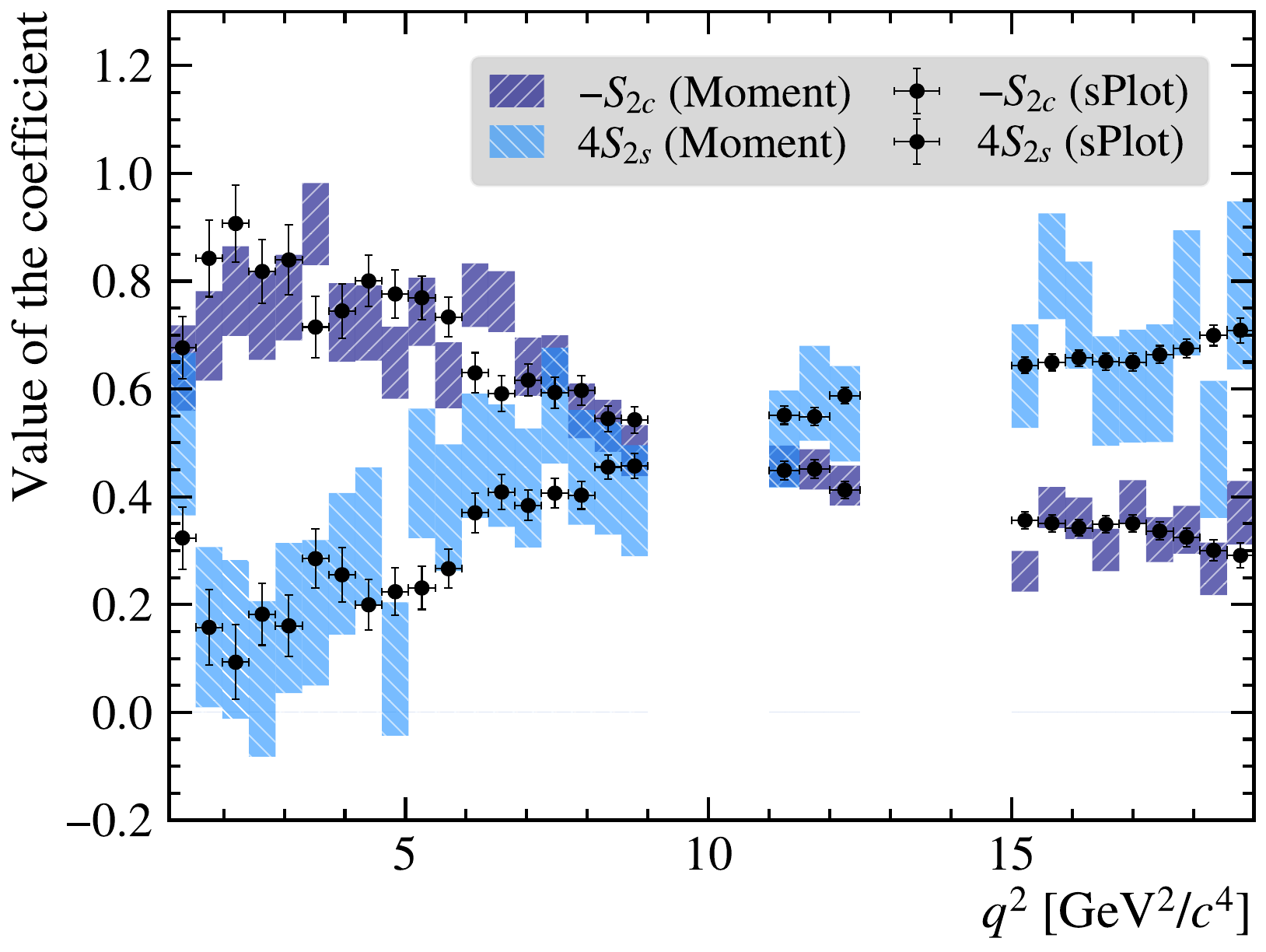}
    \caption{Values of the two angular coefficients $-S_{2s}$ and $4S_{2c}$ across the dimuon invariant-mass squared including the statistical uncertainty when extracted using the method of moments (coloured bands) and the \sPlot technique (black markers).
    All uncertainties are estimated using data sets containing 60~000 signal decays consisting of only P-wave contributions and 30~000 non-factorizing background candidates, both including non-factorizing efficiency effects.
    In the left plot, the central value is fixed to the generated value for easier comparison of the statistical uncertainties, and a standard bin width of 1 \gevgevcccc is used.
    The right plot shows an example toy without fixing the means illustrating the reduced scatter in the \sPlot method.}
    \label{fig:sweighted_unc_moments}
\end{figure}

Figure~\ref{fig:sweighted_unc_moments} shows these two angular coefficients, calculated from a mix of non-factorizing background and the purely P-wave signal sample.
The difference in shape between $-S_{2c}$ and $4S_{2s}$ and compared to the pure $n_0^\text{P}$ and $n_1^\text{P}$ shapes shown in previous figures stems from the additional normalisation to the decay rate as a function of \qsq, see Eq.\eqref{eq:angcoeffs}.
The left part of the figure uses a $1\gevgevcccc$ bin-width and fixes both results to the generated values in order to highlight the difference in uncertainty.
The uncertainties are obtained as the standard deviation of the results obtained through analysing 50 pseudo-experiments with 60~000 signal (P-wave only) and 30~000 (non-factorizing) background candidates.
The uncertainties are generally larger for the method of moments in particular at high values of \qsq.
The high-\qsq region is crucial for the investigation of the resonant contributions above the \psitwos resonance which are poorly understood and are hard to study with conventional methods.
It is known that a maximum likelihood fit results in higher precision than the method of moments.
The correlations across \qsq reduces the uncertainties even further as now every data point depends on the full data set instead of only a specific bin which constrains the scatter.
The larger amount of background compared to the signal and associated correlations contribute to generally higher uncertainties at lower \qsq for both methods.

It is important to note that the \sPlot weights calculated to separate signal from background to get the angular moments are obtained from a single fit to the four-body invariant mass.
This requires the application of the expansion trick presented in Eq.~\eqref{eq:expansion} in order properly account for the correlation between the four-body mass and the dimuon mass in the background.
Historically, moment measurements have been performed in bins of the dimuon invariant-mass squared.
On one hand, weights calculated from individual fits to the data in each bin results in larger uncertainties compared to a fit to the full data due to the reduced statistical power.
On the other hand, fitting the background in each bin individually allows to largely neglect the correlation between control and fit variables resulting in slightly more stable weights which in turn decreases the uncertainty on the weights.
Whether these two effects cancel and lead to similar overall uncertainties has not been studied in this context and likely depends strongly on the size of the considered data set, the number of data points per bin, and the expansion order used in Eq.~\eqref{eq:expansion}.

\section{Presentation of the results}
%%%Unbinned results
As already discussed in Ref.~\cite{Beck:2025qxx}, presenting an unbinned result is non-trivial.
A good solution ---though breaking the unbinned nature of this method--- is to employ histograms where central values, uncertainties, and inter-bin correlations are all well-defined.

There are different options to represent the nominal results without uncertainties in a continuous way.
The individual shapes could be represented by large unweighted data sets generated using dynamic nested sampling~\cite{Higson_2018} as implemented for example in the \textit{dynesty} package~\cite{sergey_koposov_2024_12537467} which is already widely used in the \textit{eos} flavour-fitting library~\cite{EOS}.
Another possibility is to use non-parametric models such as normalising flows.
The \textit{nabu-hep} tool~\cite{Araz:2025ezp} automatizes the modelling procedure and was specifically designed for this purpose.
Of course, the result may also be presented by publishing the data itself including the extracted weights (or an ensemble of extracted weights obtained from bootstrapping the original data) if possible and useful.

The presentation of uncertainties in a continuous way and the correct accounting for correlations across the phase space is particularly challenging.
Gaussian Processes can model mean values and uncertainty bands based on individual data points.
In our example, the data points could be obtained by producing one or more histograms of different but fixed binning for each bootstrap result.
An equivalent approach could be taken to obtain a model for the systematic uncertainty.
Finally, if all uncertainties are determined in a data-driven fashion, the weights themselves might be viewed as observables themselves and uncertainties quoted for each data point individually.

The development and study of different ideas and methods is an interesting field for future research in itself that lies outside the scope of this paper.

\section{Discussion}
%Minisummary
The foregoing discussions show that the angular coefficients $n$ in Eq.~\eqref{eq:decrate} can be extracted in an unbiased way by means of the \sPlot technique even in the presence of realistic backgrounds and complex detection efficiency.
%Uncertainties
As a direct consequence of the correlations across the phase space, the scatter around the central value is reduced compared to a measurement of the moments.
Additionally, the \sPlot method has a major advantage over per-bin likelihood fits as the width of the bins in the latter is limited by the amount of data required to achieve a stable fit.
This advantage becomes more striking when analysing larger \qsq ranges as more bins increase the complexity and bookkeeping of the binned-fit analysis while the \sPlot analysis only gains more data for the single, relatively simple, fit.
This is particularly interesting for signals with much lower production or decay rates than \BdToKpimm, including for example $\Lb\to p\Km\mumu$ or $b\to d$ transitions, where the currently available data at LHCb is barely enough to perform full angular fits in bins of \qsq~\cite{LHCb:2026dbi,LHCb:2026xvw}.

%Direct Impact
Results obtained using the presented technique would enable thorough studies of the \qsq-dependence of the different amplitudes.
What is more, better knowledge of the S-wave dihadron lineshape can reduce systematic uncertainties present in experimental measurements and help improve the understanding on $B\to M_1M_2$ form factors across the full dihadron invariant-mass range.
Another advantage over the method of moments is that this method has the ability to extract the shape of the S-wave without assuming massless leptons.
While this publication examines the \BdToKpimm decay, the derivations including the D-wave as given in App.~\ref{app:dwave} show that the method is applicable to more complex hadronic systems.
This can be particularly useful for decays such as $\Lb\to p\Km\mumu$ with lower production rates than $B$ decays but more complicated resonance spectra.

%Choices
In the presented study, we make a few simplifying assumptions that should have no impact on the validity of the results but should be mentioned regardless.
First, the efficiency is assumed to be perfectly uniform in the angle between the dihadron and dilepton decay planes, usually called $\phi$, as well as the four-body invariant-mass.
While this may not be true in real data, the efficiency weights account for any non-uniform effects or additional non-factorizing dependencies as long as they are included in the efficiency model.
Second, we assume that the signal and background efficiencies are identical.
This is not uncommon for similar measurements and requires careful systematic study to ensure negligible impact.
Third, real efficiency functions are usually much more complex than our example.
However, the key challenge are non-factorizing terms between the fit and control variables which our example proofs can be unproblematic if treated correctly.
%Similarly, background contributions can have more correlating terms.

%Drawbacks
The main disadvantage of the presented method is that it can only extract four observables when assuming only S- and P-wave contributions (six observables when including D-wave amplitudes).
Moreover, these observables are mostly amplitudes squared while interference terms vanish when integrating over $\phi$ or are inaccessible with our method as they appear with asymmetric angular functions.
In contrast, binned-fit and moment analyses can measure all angular observables, including for example the lepton-forward backward asymmetry or $P_5'$.
And amplitude fits even determine the complex phases between amplitudes directly without the need for additional interpretation.
The reader should be reminded however that deviations from the Standard Model appear in the Wilson coefficients which we access through the amplitudes.
As a consequence, the additional information provided through a measurement of the amplitudes using the \sPlot technique might help to understand the origin of the current deviations seen in amplitude combinations from conventional angular analyses.
%$A_\perp$ and $A_\parallel$ only appear together in Eq.~\eqref{eq:decrate}, this method can only provide a combined shape for these two amplitudes.

%This is easy
On a final note, we want to emphasize that this analysis method ---while new--- is in many ways simpler than conventional methods.
Like any angular analysis, the presented method relies on knowledge of the efficiency and the background shape.
However where binned-fits might suffer from small data sets in some bins and the need for a precise model of the dihadron invariant-mass, the \sPlot method can make use of the entire data set at once without any assumptions on the \mkpi lineshapes.
In comparison to amplitude analyses ---which require tremendous computational effort and assumptions when modelling the two-body invariant-masses---, the \sPlot method is almost simplistic.

\section{Conclusion}
%%%Conclusion
In summary, we showed how realistic efficiency and background can be accounted for when extracting \BdToKpimm decay amplitudes using the \sPlot method.
This paves the way for a fundamentally model-independent measurement of the shape of these amplitudes.
The dependence of the amplitudes on the dimuon invariant-mass squared is particularly interesting for the study of the underlying \bsll structure and potential deviations from the Standard Model expectation.
The shapes of the contributions to the dihadron spectrum can be used to further reduce the systematic uncertainty on experimental measurements and provide better understanding of the QCD resonance spectrum.
This method can be applied to any $B\to V(\to M_1M_2)\ellp\ellm$ decay and translated to other spin configurations after establishing an appropriate expression for the angular decay rate, which has been presented for \BdToKpimm including S-, P- and D-wave contributions.

\section*{Acknowledgements}
We are grateful to Leon Carus for providing the simulation samples used in this study.
We want to appreciate useful discussions with Christoph Langenbruch.
We would like to thank the US National Science Foundation, whose funding under award numbers 2310073 and 2607794 have helped support this work.
T.X. wants to acknowledge support obtained from ETH Zurich through the Swiss-European Mobility Programme Worlwide Project \& Traineeships.

We want to acknowledge the usage of generative AI tools in the three limited and specific ways outlined in the following.
First, AI was employed as a research and reference tool to locate relevant literature and resources when standard search methods proved insufficient, mainly serving as a sophisticated memory aid to clarify concepts and ensure a complete literature review.
Second, AI was used to assist with code debugging and refinement.
However, all code was rigorously tested and held to the same high standards regardless of whether it was written with or without AI assistance.
Third, during the textual review before submission to a journal, the wording in individual sentences was improved using AI suggestions.
In all instances, the authors reviewed and edited all AI-generated suggestions and take full responsibility for the accuracy and integrity of the final work.

\appendix
\section{Extension to a decay rate including D-wave}\label{app:dwave}
\subsection{Structure of the angular decay rate}
In the presence of massive leptons and scalar amplitudes, the two-dimensional angular decay rate of \BdToKpimm including S-, P-, and D-wave contributions can be expressed as
\begin{align}\label{eq:dwave_decrate}
\begin{split}
\frac{1}{\Gamma_\text{total}}\frac{\deriv\Gamma}{\deriv\cos\th\deriv\cos\tl} 
    &=f_{\text{S0}}^h \left[n_0^\text{S} f_-^\ell+b_0^\text{S}-f_1^\ell a_{s0}^\text{S}\right] \\
    &+f_{\text{P0}}^h \left[n_0^\text{P} f_-^\ell+b_0^\text{P}-f_1^\ell a_{s0}^\text{P}\right] \\
    &+f_{\text{D0}}^h \left[n_0^\text{D} f_-^\ell+b_0^\text{D}-f_1^\ell a_{s0}^\text{D}\right] \\
    &+f_{\text{DS}}^h \left[n_0^{\text{DS}} f_-^\ell+b_0^\text{DS}-f_1^\ell a_{s0}^{\text{DS}}\right] \\
    &+f_{\text{P1}}^h \left[n_1^\text{P} f_+^\ell+b_1^\text{P}-2 a_1^\text{P} f_1^\ell\right] \\
    &+f_{\text{D1}}^h \left[n_1^\text{D} f_+^\ell+b_1^\text{D}-2 a_1^\text{D} f_1^\ell\right] \\
    &+f_{\text{DP0}}^h \left[a_0^{\text{DP}} f_-^\ell+b_0^\text{DP}-f_1^\ell a_{\text{s0}}^{\text{DP}}\right] \\
    &+f_{\text{SP0}}^h \left[a_0^{\text{PS}} f_-^\ell+b_0^\text{PS}-f_1^\ell a_{s0}^{\text{PS}}\right] \\
    &+f_{\text{DP1}}^h \left[\frac{1}{2} a_{\text{same}}^{\text{DP}} f_+^\ell+b_1^\text{DP}+f_1^\ell a_{\text{opp}}^{\text{DP}}\right] \ . \\
\end{split}
\end{align}%
The angular functions $f^h$ ($f^\ell$) depend implicitly on the hadron (lepton) helicity angle and the coefficients $n$, $b$, and $a$ depend on the dihadron and dilepton invariant-masses.
The symmetric angular functions with hadron-dependence are
\begin{align}\label{eq:symmetrichadron}
\begin{split}
    f_\text{S0}^h(\th) &= \frac{3}{8} \ , \\
    f_\text{P0}^h(\th) &= \frac{9}{8}\cos^2\th \ , \\
    f_\text{P1}^h(\th) &= \frac{9}{32}(1-\cos^2\th) \ , \\
    f_\text{D0}^h(\th) &= \frac{15}{32}(1-3\cos^2\th)^2 \ , \\
    f_\text{D1}^h(\th) &= \frac{45}{32}\cos^2\th(1-\cos^2\th) \ , \\
    f_\text{DS}^h(\th) &= -\frac{3}{16}\sqrt{5}(1-3\cos^2\th) \ , \\
\end{split}
\end{align}
where the lower index indicates which partial waves are affected.
All of those functions are strictly non-negative because they come with the absolute values squared of individual amplitudes except $f_\text{DS}^h$ which is accompanied by the D-/S-wave interference terms and whose integral vanishes.
The asymmetric angular functions with hadron-dependence are
\begin{align}
\begin{split}
    f_\text{DP0}^h(\th) &= -\frac{3}{16}\cos\th(1-3\cos^2\th) \ , \\
    f_\text{SP0}^h(\th) &= \frac{3}{8}\sqrt{3}\cos\th \ , \\
    f_\text{DP1}^h(\th) &= \frac{9}{16}\sqrt{5}\cos\th(1-\cos^2\th) \ , \\
\end{split}
\end{align}
corresponding to the interference of P-wave amplitudes with D- or S-wave amplitudes.
Note that the interference between the D- and S-wave amplitudes is symmetric while the interferences with the P-wave amplitudes are asymmetric because the S- and P-wave are assumed to have the same parity different from the P-wave.
The angular functions with lepton-dependence are
\begin{align}
\begin{split}
    f^\ell_1(\tl) = \cos\tl \qquad\text{and}\qquad f_\pm^\ell(\tl) = 1\pm\cos^2\tl \ .
\end{split}
\end{align}

For all coefficients, $a$, $b$, and $n$, the upper index indicates which partial wave contributes (S, P, D) and the lower index indicates the nature of the dilepton system in the two interfering amplitudes (scalar $s$, time-like $t$, longitudinal $0$, transverse $1$).
The coefficients $a$ correspond to interference terms between different amplitudes and appear with asymmetric angular terms.
As a consequence, they vanish when integrating over the angles to determine the normalization condition:
\begin{align}
    1 = n^\text{D}_0+n^\text{D}_1+n^\text{P}_0+n^\text{P}_1+n^\text{S}_0 + \frac{3}{4}(b^\text{D}_1 + b^\text{D}_0 + b^\text{P}_1 + b^\text{P}_0 + b^\text{S}_0) \ .
\end{align}
The coefficients $b$ are combinations of different contributions and vanish in the SM ($A_s=0$) with massless leptons ($\beta\to1$):
\begin{align}\label{eq:bterms-dwave}
\begin{split}
b_0^\text{S} &= \frac{1}{2} \left(\frac{1-\beta^2}{\beta^2}\left(a_{\text{LR0}}^\text{S}+n_t^\text{S}+n_0^\text{S}\right)+n_s^\text{S}\right) \ , \\
b_0^\text{P} &= \frac{1}{2} \left(\frac{1-\beta^2}{\beta^2}\left(a_{\text{LR0}}^\text{P}+n_t^\text{P}+n_0^\text{P}\right)+n_s^\text{P}\right) \ , \\
b_0^\text{D} &= \frac{1}{2} \left(\frac{1-\beta^2}{\beta^2}\left(a_{\text{LR0}}^\text{D}+n_t^\text{D}+n_0^\text{D}\right)+n_s^\text{D}\right) \ , \\
b_1^\text{P} &= \frac{1-\beta^2}{\beta^2}\left(a_{\text{LR1}}^\text{P}+n_1^\text{P}\right) \ , \\
b_1^\text{D} &= \frac{1-\beta^2}{\beta^2}\left(a_{\text{LR1}}^\text{D}+n_1^\text{D}\right) \ , \\
b_1^\text{DP} &= \frac{1}{2} \frac{1-\beta^2}{\beta^2}\left(a_{\text{LR1}}^{\text{DP}}+a_{\text{same}}^{\text{DP}}\right) \ , \\
b_0^\text{DS} &= \frac{1}{2} \left(\frac{1-\beta^2}{\beta^2}\left(a_{\text{LR0}}^{\text{DS}}+a_t^{\text{DS}}+a_0^{\text{DS}}\right)+a_s^{\text{DS}}\right) \ , \\
b_0^{SP} &= \frac{1}{2} \left(\frac{1-\beta^2}{\beta^2}\left(a_{\text{LR0}}^{\text{PS}}+a_t^{\text{PS}}+a_0^{\text{PS}}\right)+a_s^{\text{PS}}\right) \ , \\
b_0^\text{DP} &= \frac{1}{2} \left(\frac{1-\beta^2}{\beta^2}\left(a_{\text{LR0}}^{\text{DP}}+a_t^{\text{DP}}+a_0^{\text{DP}}\right)+a_s^{\text{DP}}\right) \ . \\
\end{split}
\end{align}
\subsection{Expressions for the angular coefficients}
In the following, we give the dependence of the coefficients on the transversity amplitudes.
Because each partial wave contributes the same amplitude combinations, we express the combinations using generic amplitudes $A$ and $B$, which can stand for S, P, D where the lower index is $s,t,0$ and P or D if the lower index is $1$ or $\perp,\parallel$.

The $n$ terms represent the amplitudes squared of each partial wave
\begin{align}
\begin{split}
n_i^A &=  \beta ^2 |A_i| ^2 \ , \quad i=s,t \ , \\
n_0^A &=  \beta ^2 \left(| A_0^L| ^2 + | A_0^R| ^2\right) \ , \\
n_1^A &=  \beta ^2 \left(| A_\parallel^L| ^2+| A_\perp^L| ^2+| A_\parallel^R| ^2+| A_\perp^R| ^2\right) \ , \\
\end{split}
\end{align}
where $n_1^A$ only appears for the P- and D-wave.
Each partial wave has terms corresponding to the interference between different amplitudes,
\begin{align}
\begin{split}
    a_{s0}^A &= 2 \frac{\beta  m_\ell }{\sqrt{\qsq}}\Re\left[A_s (A_0^L+A_0^R)^*\right] \ , \\
    a_{LR0}^A &= 2 \beta ^2 \Re\left[A_0^L (A_0^R)^*\right] \ , \\
    a_{LR1}^A &= 2 \beta ^2 \Re\left[A_\parallel^L (A_\parallel^R)^*+A_\perp^L (A_\perp^R)^*\right] \ , \\
    a_1^A &= 2 \beta  \Re\left[-A_\parallel^L (A_\perp^L)^*+A_\parallel^R (A_\perp^R)^*\right] \ . \\
\end{split}
\end{align}
As before, the coefficients $a_{LR1}^A$ and $a_1^A$, which include $A_{\perp,\parallel}$ amplitudes, only appear for the P- and D-wave.
Note that $a_1^A$ is the lepton-side forward-backward asymmetry.

There are a number of interaction terms between the scalar and longitudinal amplitudes between different states,
\begin{align}
\begin{split}
    a_t^{AB} &= 2 \beta ^2 \Re\left[A_t B_t^*\right] \ , \\
    a_s^{AB} &= 2 \beta ^2 \Re\left[A_s B_s^*\right] \ , \\
    a_0^{AB} &= 2 \beta ^2 \Re\left[A_0^L (B_0^L)^*+A_0^R (B_0^R)^*\right] \ , \\
    a_{LR0}^{AB} &= 2 \beta ^2 \Re\left[A_0^L (B_0^R)^*+B_0^L (A_0^R)^*\right] \ , \\
    a_{s0}^{AB} &= 2 \frac{\beta  m_\ell}{\sqrt{\qsq}} \Re\left[A_s (B_0^L+B_0^R)^*+B_s (A_0^L+A_0^R)^*\right] \ , \\
\end{split}
\end{align}
where $AB$ can be any of the combinations SP, SD, or DP.
Finally, there are interaction terms between the amplitudes with dilepton helicity $\pm1$, which only exist between two states which both have spin $J\geq1$, in this case the combination DP,
\begin{align}
\begin{split}
a_{\text{same}}^\text{DP} &= 2 \beta ^2 \Re\left[P_\parallel^L (D_\parallel^L)^*+P_\perp^L (D_\perp^L)^*+P_\parallel^R (D_\parallel^R)^*+P_\perp^R (D_\perp^R)^*\right] \ , \\
a_{\text{opp}}^\text{DP} &= 2 \beta  \Re\left[P_\parallel^L (D_\perp^L)^*+P_\perp^L (D_\parallel^L)^*-P_\parallel^R (D_\perp^R)^*-P_\perp^R (D_\parallel^R)^*\right] \ , \\
a_{LR1}^\text{DP} &= 2 \beta ^2 \Re\left[P_\parallel^L (D_\parallel^R)^*+P_\parallel^R (D_\parallel^L)^*+P_\perp^L (D_\perp^R)^*+P_\perp^R (D_\perp^L)^*\right] \ . \\
\end{split}
\end{align}
\subsection{\sPlot formalism}
In order to apply the \sPlot formalism to the decay rate given in Eq.~\eqref{eq:dwave_decrate}, it needs to be expressed using a basis of linearly independent angular terms.
A possible basis is for example the basis used to extract angular moments, see e.g. Ref.~\cite{Gratrex:2015hna}.
Another convenient option is
\begin{align}\label{eq:splot-dwave}
\begin{split}
    \frac{1}{\Gamma_\text{total}}\frac{\deriv\Gamma}{\deriv\cos\th\deriv\cos\tl}
    &=f_\text{S0}^h f_{-}^\ell\left[n_0^\text{S}-\frac{\sqrt{5}}{4}\left(2n_0^{DS}+b_0^{DS}\right)+\frac{3}{8}b_1^\text{P}+\frac{5}{24}b_1^\text{D} +\frac{1}{2}b_0^\text{S}\right] \\
    &+f_\text{P0}^h f_{-}^\ell\left[n_0^\text{P} + \frac{\sqrt{5}}{4}\left(2n_0^{DS} + b_0^{DS}\right)-\frac{1}{8}b_1^\text{P}+\frac{5}{24}b_1^\text{D}+\frac{1}{2}b_0^\text{P}\right] \\
    &+f_\text{D0}^h f_{-}^\ell\left[n_0^\text{D}-\frac{1}{6}\left(b_1^\text{D}-3b_0^\text{D}\right)\right] \\
    &+\frac{3}{16}f_{+}^\ell \left[5b_0^\text{D}+3b_0^\text{P}+b_0^\text{S} + \sqrt{5}b_0^{DS}\right] \\
    &+f_\text{P1}^h f_{+}^\ell \left[n_1^\text{P}+\frac{1}{2}b_1^\text{P}-\frac{5}{2}b_0^\text{D}-2b_0^\text{P} - \sqrt{5}b_0^{DS}\right] \\
    &+f_\text{D1}^h f_{+}^\ell \left[n_1^\text{D}+\frac{1}{2}\left(b_1^\text{D}-3b_0^\text{D}\right)\right] \\
    &+f_1^\ell\left(c_{10}+c_{11}\cos\tl+c_{12}\cos^2\tl+c_{13}\cos^3\th+c_{14}\cos^4\th\right) \\
    &+\cos\th\left(c_{1+}f_+^\ell+c_{1-}f_-^\ell\right) \\
    &+\cos^3\th\left(c_{3+}f_+^\ell+c_{3-}f_-^\ell\right) \ . \\
\end{split}
\end{align}
The angular space only allows for six independent symmetric angular functions, represented by the first six lines.
In the following, the six extractable shapes accompanying these functions are called $(n_0^\text{S})'$, $(n_0^\text{P})'$, $(n_0^\text{D})'$,  $(n_\beta)'$, $(n_1^\text{P})'$, and $(n_1^\text{D})'$ representing the expressions in square brackets in order of appearance.
Note that the terms containing interference terms of the D and S wave amplitudes with longitudinal dimuon polarization have to be absorbed into the terms coming with the functions $f_{S0}^h$ and $f_{P0}^h$ as $f_{DS}^h$ is a linear combination of the two.
All coefficients appearing with asymmetric angular terms are combined into the coefficients $c$:
\begin{align}
\begin{split}
    c_{10} &= \frac{3}{32} \left(-5a_{s0}^\text{D}+2\sqrt{5} a_{s0}^{\text{DS}}-6a_1^\text{P}-4 a_{s0}^\text{S}\right) \\
    c_{11} &= \frac{3\sqrt{3} }{16} \left(\sqrt{15} a_{\text{opp}}^{\text{DP}}+\sqrt{5} a_{s0}^{\text{DP}}-2 a_{s0}^{\text{PS}}\right) \ , \\
    c_{12} &= \frac{9}{16} \left(-5 a_1^\text{D}+5 a_{s0}^\text{D}- \sqrt{5} a_{s0}^{\text{DS}}-2a_{s0}^\text{P}+a_1^\text{P} \right) \ , \\
    c_{13} &= -\frac{9}{16} \sqrt{5} \left(a_{\text{opp}}^{\text{DP}}+\sqrt{3} a_{s0}^{\text{DP}}\right) \ , \\
    c_{14} &= \frac{45}{32}\left(2 a_1^\text{D}-3 a_{s0}^\text{D}\right) \\
    c_{1+} &= +\frac{3\sqrt{3}}{32}\left(-\sqrt{5} b_0^\text{DP}+\sqrt{15}b_1^\text{DP}+2b_0^\text{PS}+\sqrt{15} a_{\text{same}}^{\text{DP}} \right) \ , \\
    c_{1-} &= +\frac{3\sqrt{3}}{32}\left(-\sqrt{5} b_0^\text{DP}+\sqrt{15} b_1^\text{DP}+2 b_0^\text{PS}-2\sqrt{5} a_0^{\text{DP}}+4a_0^{\text{PS}}\right) \ , \\
    c_{3+} &= -\frac{9\sqrt{5}}{32}\left(-\sqrt{3} b_0^\text{DP}+ b_1^\text{DP}+ a_{\text{same}}^{\text{DP}}\right) \ , \\
    c_{3-} &= -\frac{9\sqrt{5}}{32}\left(-\sqrt{3} b_0^\text{DP}+ b_1^\text{DP}-2 \sqrt{3} a_0^{\text{DP}}\right) \ . \\
\end{split}
\end{align}

In order to illustrate the possibilities of amplitude extraction in this scenario, a simple toy data set is generated.
In this model, the dependence on the dihadron mass is modelled using one Gaussian for each of the $K(800), K(892), K_0(1430), K_2(1430)$ states and the dependence on the dimuon invariant-mass squared are different polynomials multiplied with $\beta^2$ or $\tfrac{1-\beta^2}{\beta^2}$ and a factor forcing the amplitudes to zero at the phase space boundaries.
The magnitude of the components is inspired by an amplitude analysis of $\Bd\to \Kp\pim\jpsi$~\cite{Belle:2014nuw}.
The magnitude of the $b$ terms, see Eq.~\eqref{eq:bterms-dwave}, is set to 2\% of the corresponding $n$ coefficient.
Because the asymmetry terms $c$ do not affect the extraction of the amplitudes, they are all neglected in this study.

\begin{figure}[t]
    \centering
    \includegraphics[width=0.5\linewidth]{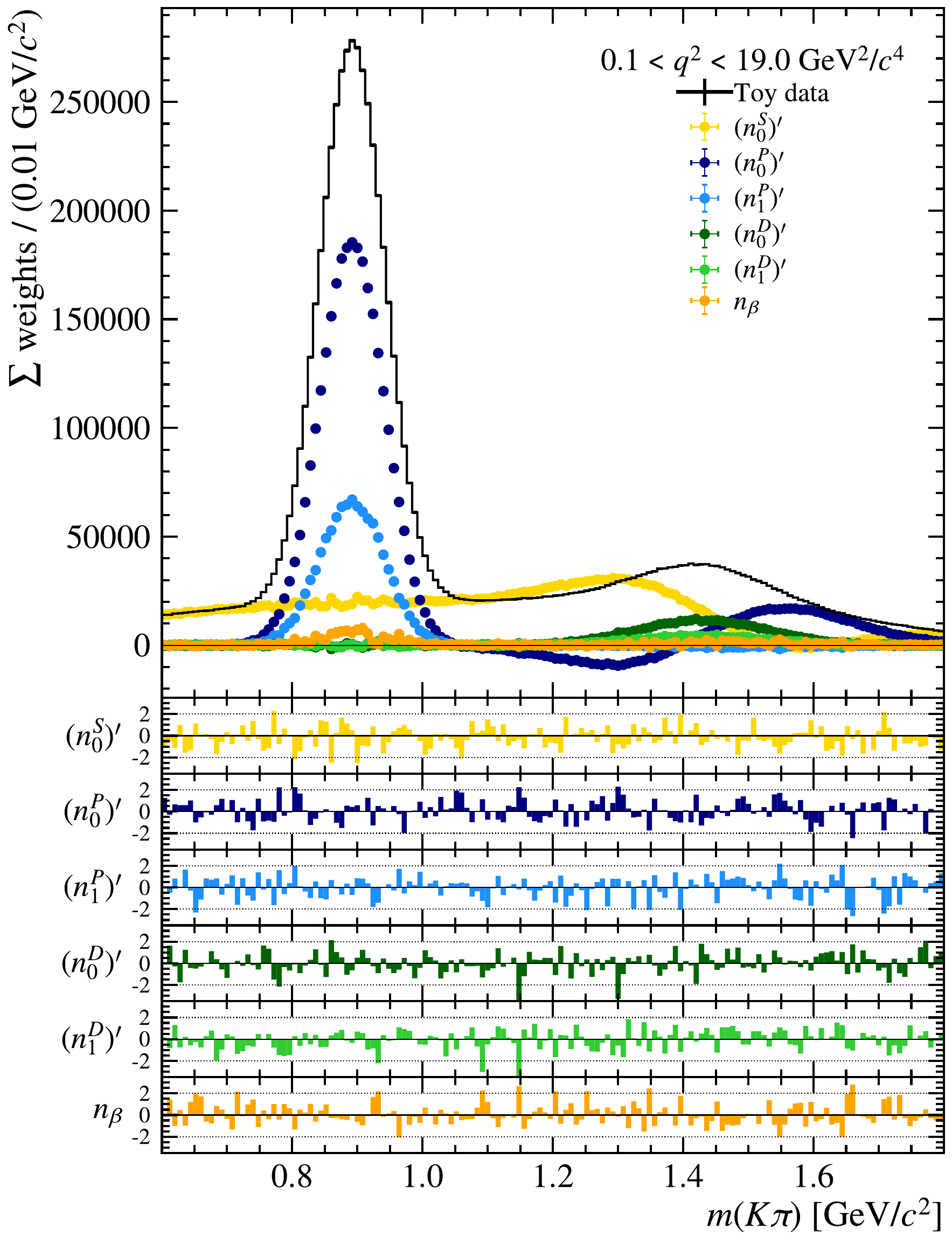}%
    \includegraphics[width=0.5\linewidth]{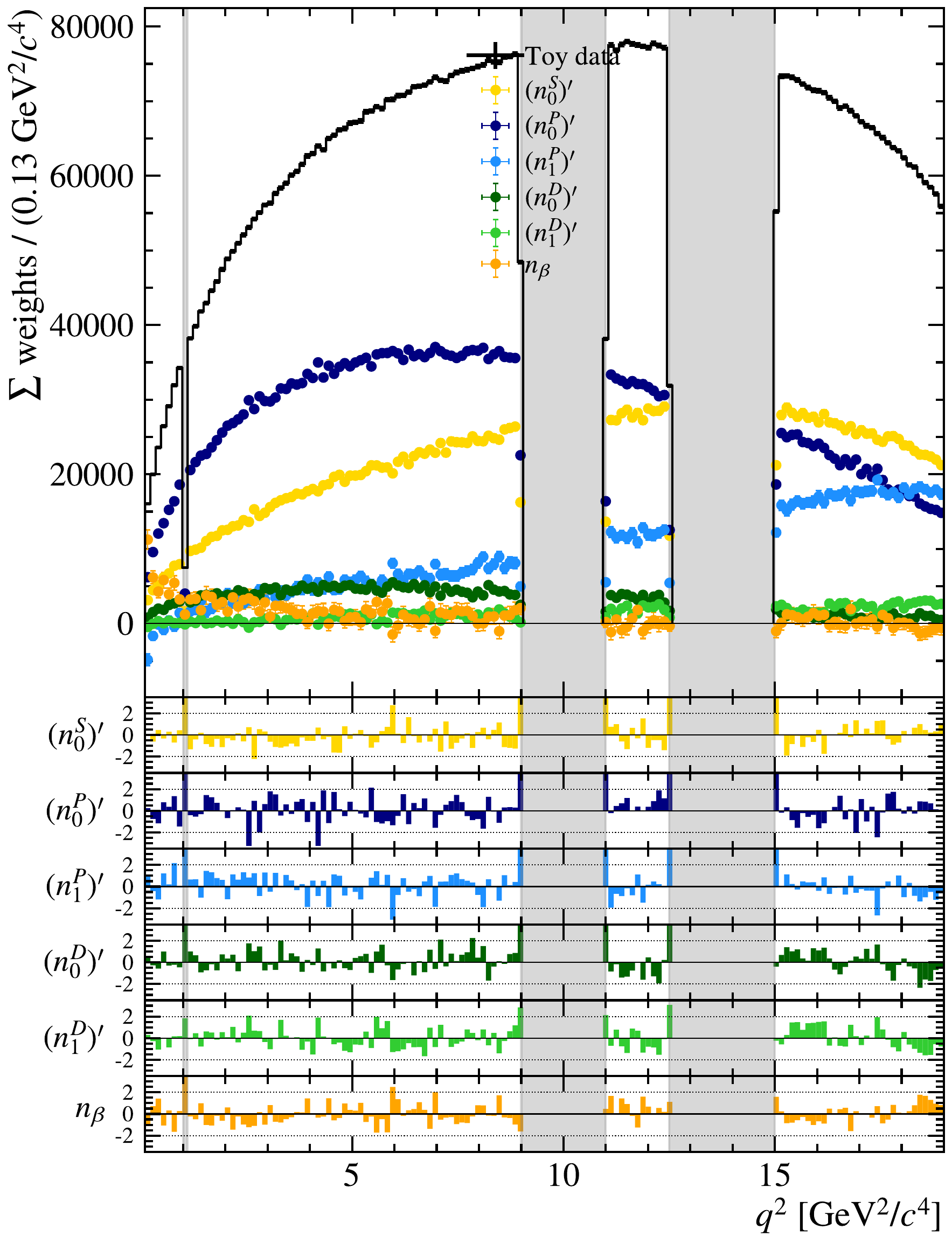}
    \caption{Extracted $(n)'$ shapes in the (left) dihadron invariant-mass and (right) dimuon invariant-mass squared from a large toy data set including S-, P-, and D-wave components including (bottom parts) pulls with respect to the true shape.}
    \label{fig:dwave-primed}
\end{figure}

After a fit to a very large toy sample where the $(n)'$ coefficients are floating, \sPlot weights can be calculated as described in the main part of this paper.
Figure~\ref{fig:dwave-primed} shows that these shapes can be extracted without bias.

\begin{figure}
    \centering
    \includegraphics[width=0.5\linewidth]{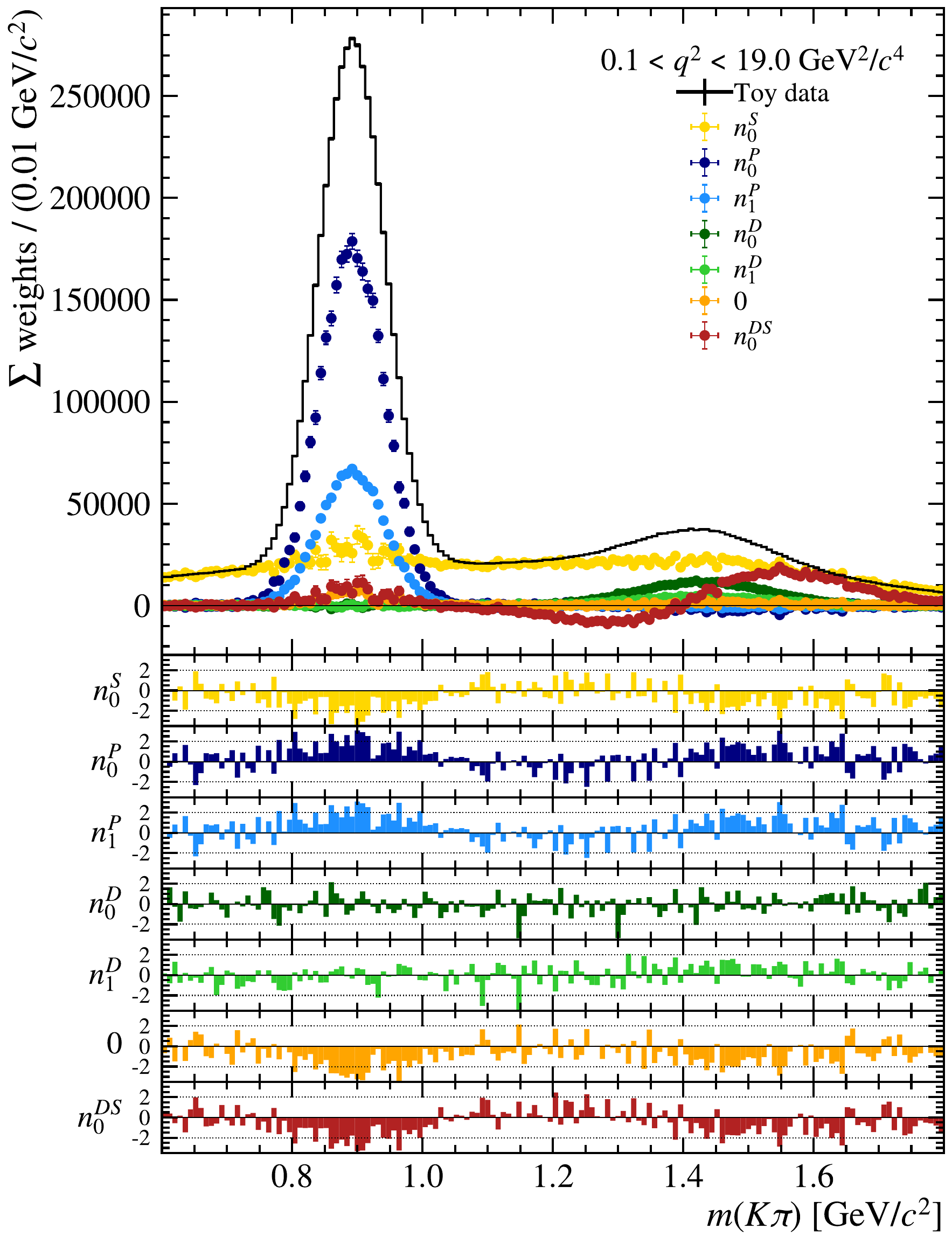}%
    \includegraphics[width=0.5\linewidth]{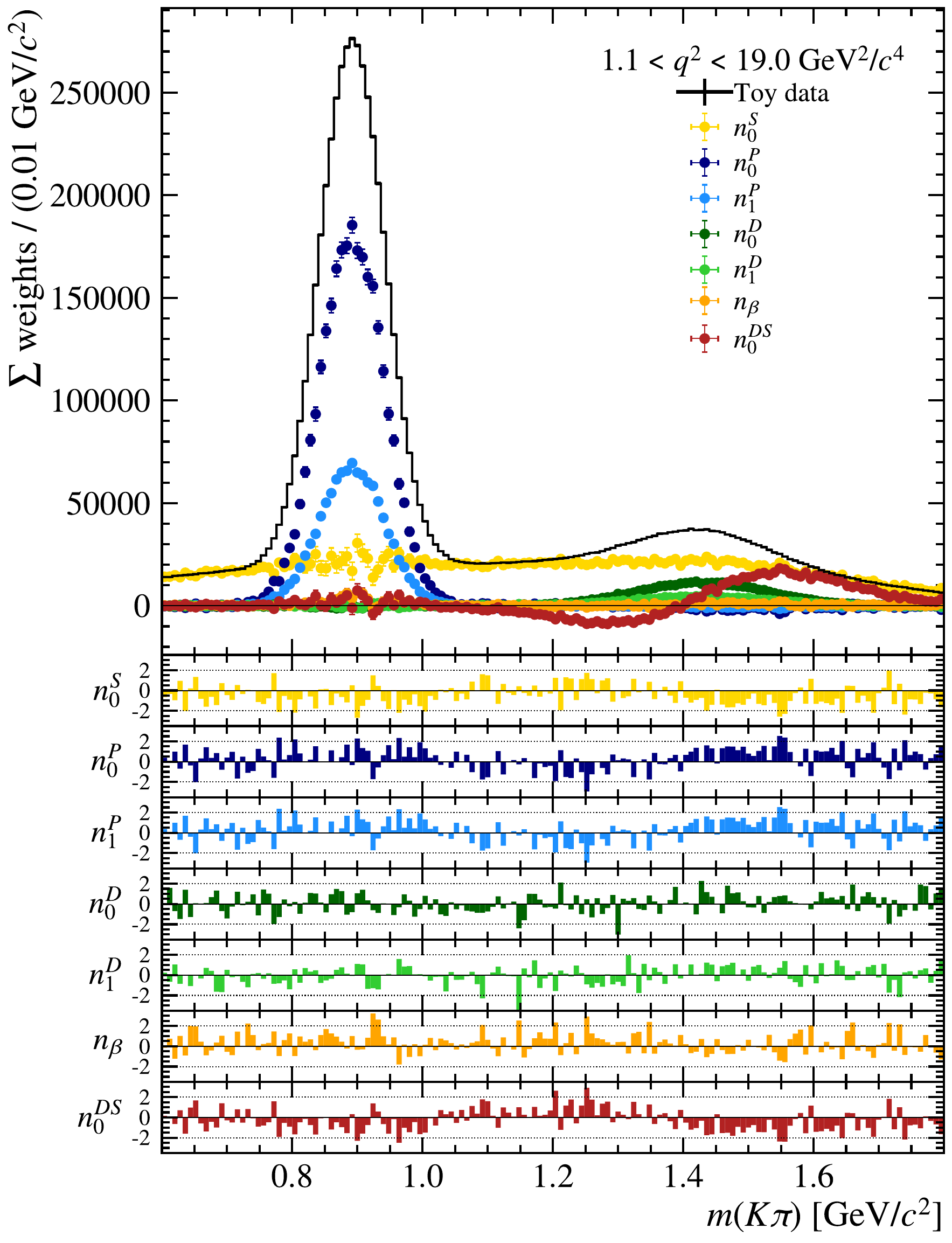}
    \caption{Extracted $n$ shapes in the dihadron invariant-mass from a large toy data set including S-, P-, and D-wave components in the (left) full \qsq range and (right) removing the low \qsq region.
    The DS interference and the $n_0^\text{S,P}$ terms are disentangled by assuming the two P-wave contributions have the same shape.
    The pull plots compare the extracted shape to the $n$ shapes neglecting $b$-terms.
    Note that the bias is only visible in a very large data set and exaggerated in this comparison as all $b$-terms contribute to the bias while only $b$-terms of different partial waves would bias the shape.}
    \label{fig:mkpi-unprimed}
\end{figure}

As a final result, we want to show that the dependency of the $n$ coefficients on \mkpi, i.e. the pure S-, P-, and D-wave shapes without contaminations, can still be extracted even in this complex setup.
In a region where the dimuon mass is negligible, $m_l\ll\qsq$, the $b$-terms vanish which allows to directly access the P-wave shape via $(n_1^\text{P})'=n_1^\text{P}$ and the D-wave shape via $(n_1^\text{D})'=n_1^\text{D}$ or $(n_0^\text{D})'=n_0^\text{D}$.
One can assume that all pure P-wave amplitudes have the same shape in the dihadron invariant-mass.
This means that in the absence of the $b$-terms, the weights, $w$, extracting the shape of $n_1^\text{P}$ in the dihadron invariant-mass can be scaled to extract the shape of $n_0^\text{P}$,
\begin{align}
    w_0^\text{P} = \frac{n_0^\text{P}}{n_1^\text{P}} (w_1^\text{P})' \overset{\beta\to1}{\approx} \frac{n_0^\text{P}}{n_1^\text{P}} w_1^\text{P} \ .
\end{align}
This trick requires reliable knowledge about the relative size of the two $P$-wave contributions, $\tfrac{n_0^\text{P}}{n_1^\text{P}}$, which could be obtained for example from a fit to the angular distribution in a narrow window around the \Kstarz peak where the D-wave ---and by construction this includes the DS interference--- is negligible.
As a result, the shape of the pure S-wave contribution in \mkpi can be extracted by recombining the \sPlot weights calculated for the $(n_0^\text{S})'$, $(n_0^\text{P})'$, and $n_1^\text{P}$ contributions,
\begin{align}
    (w_0^\text{S})' + (w_0^\text{P})' - \tfrac{n_0^\text{P}}{n_1^\text{P}} (w_1^\text{P})' \overset{\beta\to1}{\approx} n_0^\text{S}+ n_0^\text{P} - \tfrac{n_0^\text{P}}{n_1^\text{P}} w_1^\text{P} = w_0^\text{S} \ .
\end{align}
Figure~\ref{fig:mkpi-unprimed} illustrates the possibilities of this approximation when (left) considering $0.1\gevgevcccc<\qsq$ or (right) $1.1\gevgevcccc<\qsq$.
The top parts of the plots show the shapes extracted from the toy data set and the pull plots compare the extracted shapes to the individual $n$ terms while neglecting the $b$-terms.
The inclusion of low \qsq values in the left figure results in small biases visible in very large data sets.
Note that these biases are due to the $b$-terms present in the weighted distributions but not in the reference.
As mentioned previously, all amplitudes of a given partial wave likely share the same \mkpi dependence.
This means that the perceived bias here is overrepresented as $b_0^\text{P}$ and $b_1^\text{P}$ for example add to the bias in the $n_0^\text{P}$ magnitude while they actually have the exact same shape and would hence not add to a shape bias.
The right part of the figure only uses \qsq above 1.1\gevgevcccc in the fit and when calculating the \sPlot weights which reduces the bias to negligible levels in this setup.
As discussed before and in the main part of this paper, the shown bias in the full range is overstated and negligible in a data set of realistic size.

\clearpage
\section{Efficiency shape}\label{app:eff}
\begin{figure}[!ht]
    \centering
    \includegraphics[width=\linewidth]{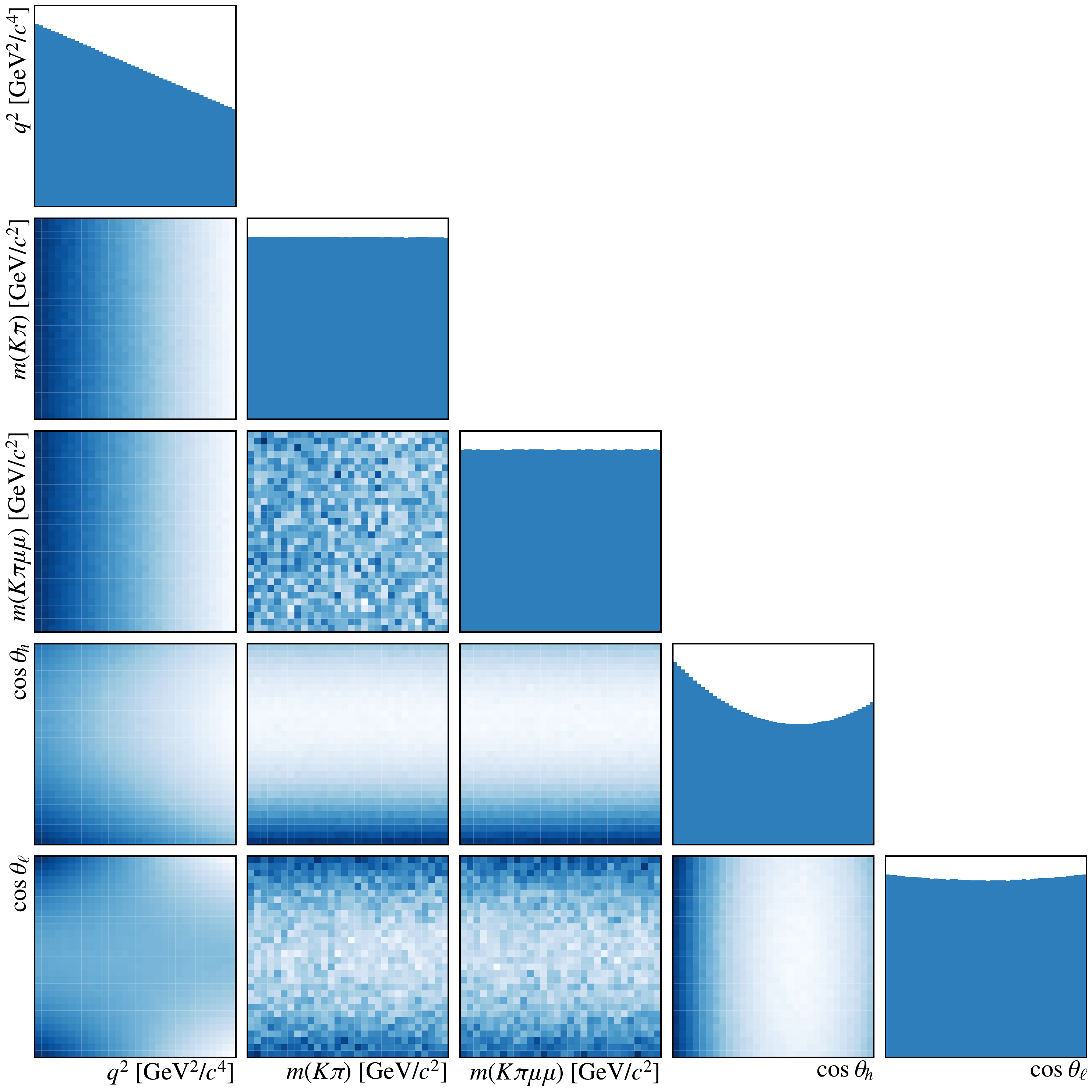}
    \caption{One- and two-dimensional projections of the efficiency shape.}
    \label{fig:efficiency}
\end{figure}

Fig.~\ref{fig:efficiency} shows the shape of the efficiency in one- and two-dimension projections onto the five-dimensional space considered in the analysis.
We assume a uniform efficiency for the four-body invariant-mass and a small linear dependence on \mkpi and \qsq.
The efficiency projected onto the cosine of either helicity angle peaks at zero and drops towards $\pm1$ with a stronger drop in the muon angle, $\cos\tl$, due to their much smaller mass than protons or kaons.
If the two hadrons are different, as is the case for \BdToKpimm, the efficiency is asymmetric in the cosine of the hadron helicity angle, $\cos\th$.
The efficiency shape for the lepton helicity angle varies across the dimuon invariant-mass squared.
Due to four-momentum conservation, candidates with low dimuon invariant mass, tend to have large dimuon momentum in the four-body rest-frame.
If this momentum aligns with the individual muon momentum in the dimuon rest frame, i.e. $\cos\tl\approx\pm1$, the muons can have very different momenta after a boost into the lab frame.
As a consequence, the efficiency at low dimuon invariant-mass and $\cos\tl\approx\pm1$ tends to drop to zero because the lower momentum muon does not pass data quality selections.
This effect becomes less relevant for larger dimuon invariant-mass because the boost into the four-body rest frame creates less asymmetry between the muon momenta.

\clearpage
\section{Background distributions}\label{app:bkg}
\begin{figure}[!ht]
    \centering
    \includegraphics[width=\linewidth]{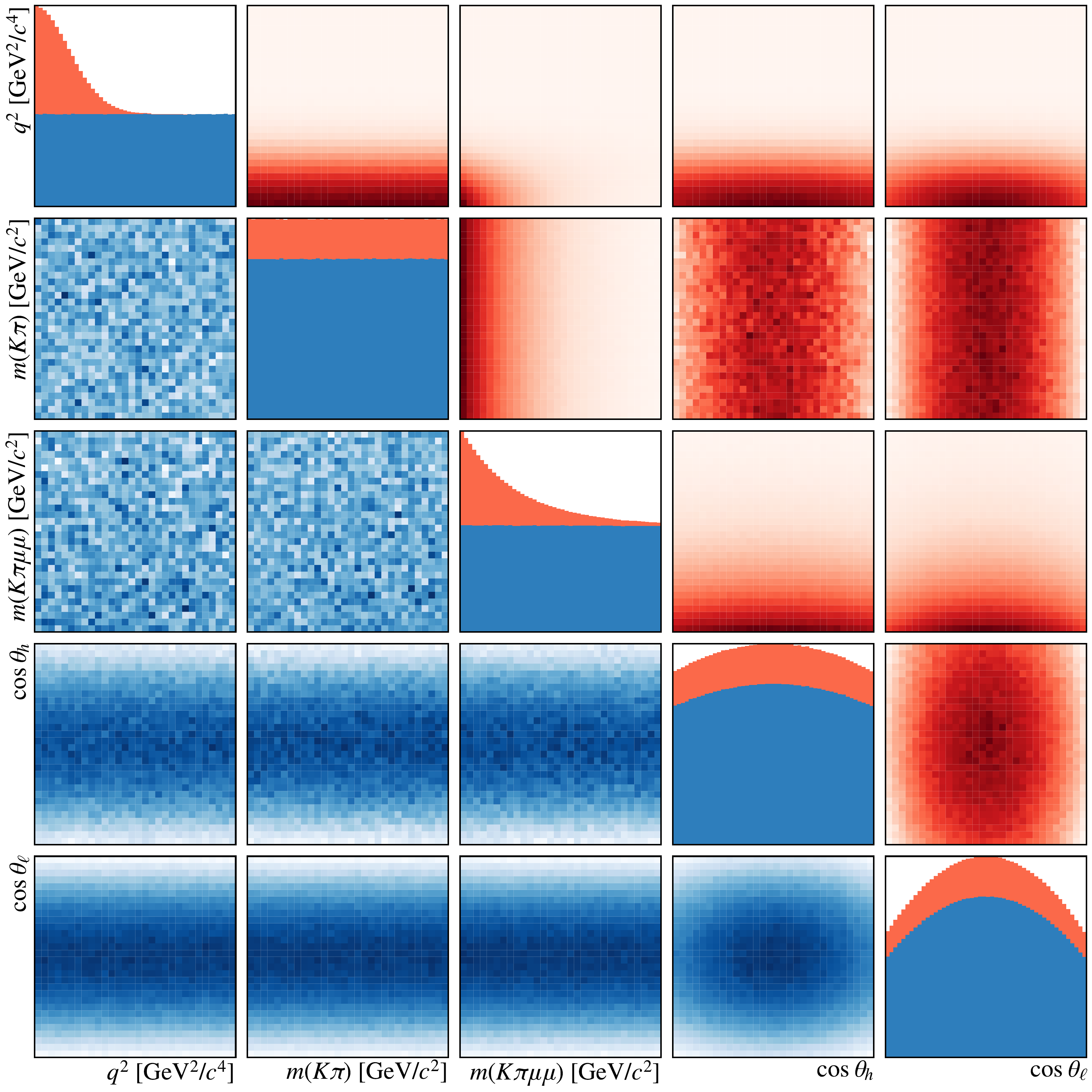}
    \caption{One- and two-dimensional projections of the background samples. The combinatorial background sample without correlations and non-factorizing terms is plotted in the lower left subplots and in blue. The sample introducing correlations between the four-body invariant-mass and the dimuon invariant-mass squared is plotted in the upper right subplots and in red.}
    \label{fig:background}
\end{figure}
\clearpage

\section{Figures for massive muons}\label{app:massive_muons}

\begin{figure}[!ht]
    \centering
    \includegraphics[width=.49\textwidth]{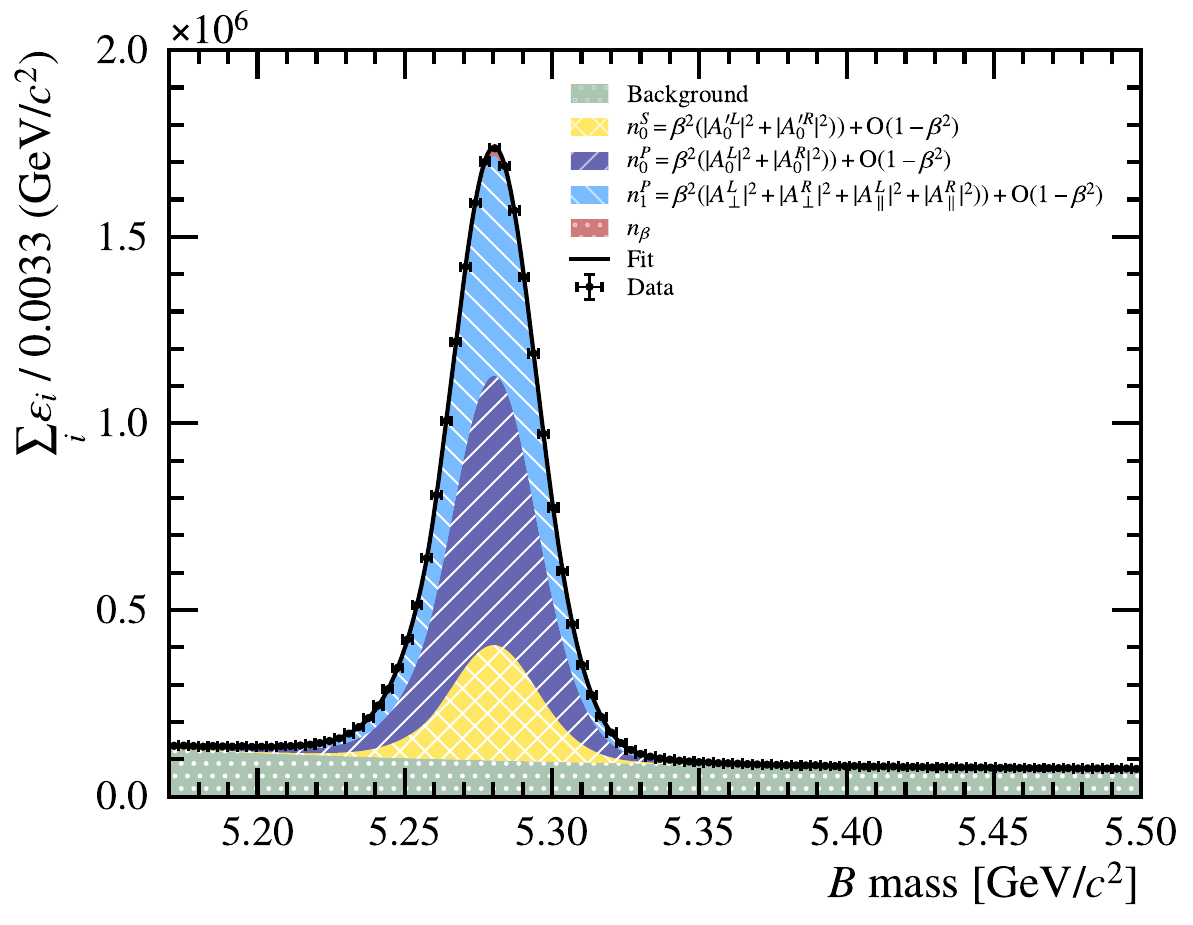}
    \includegraphics[width=.49\textwidth]{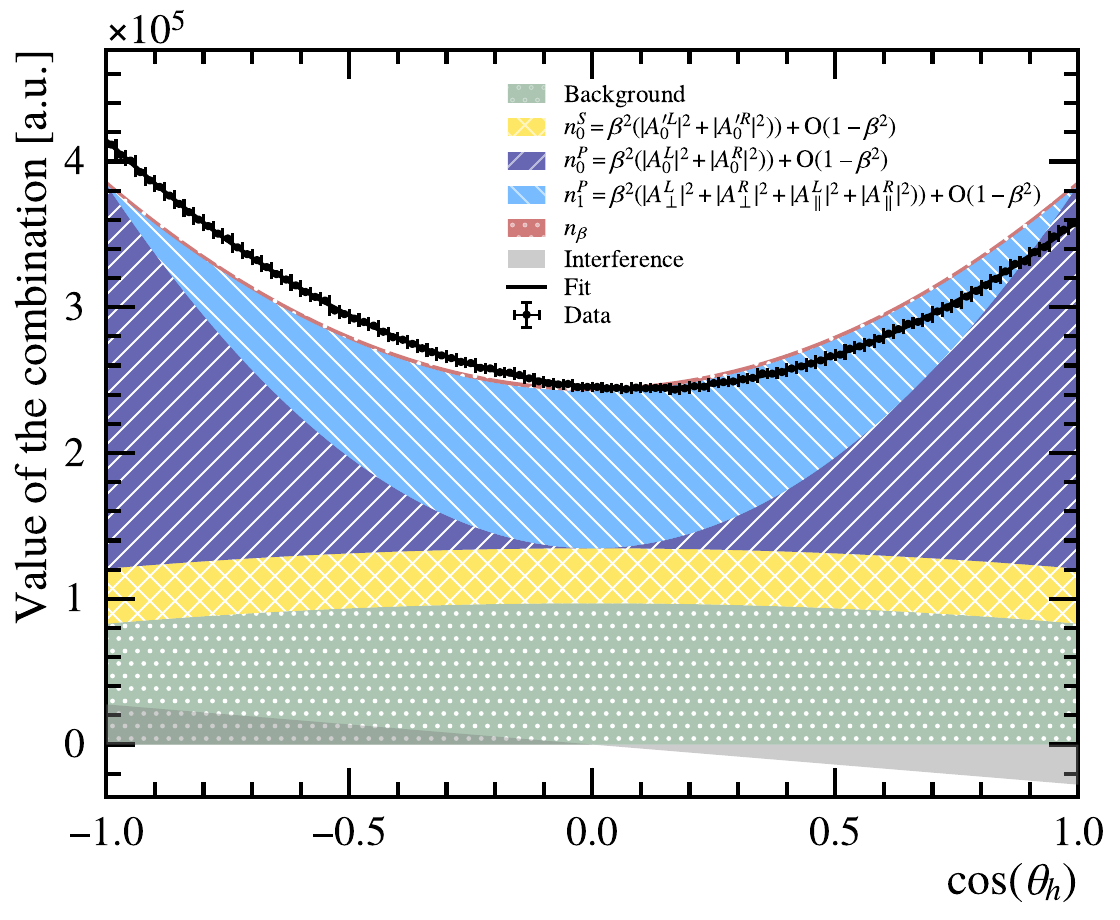}%
    \includegraphics[width=.49\textwidth]{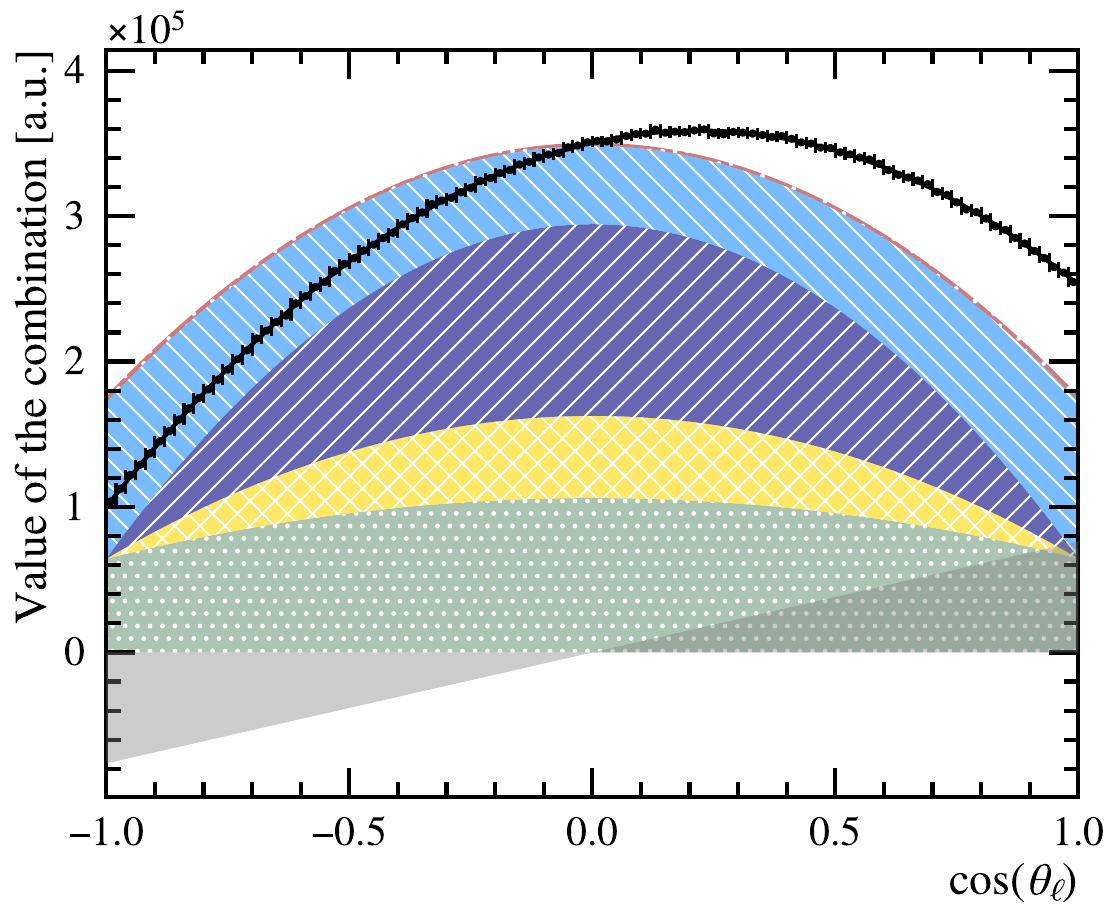}
    \caption{Projection of a large toy data set of more than 10 million data points, generated assuming massive muons, onto the (top) four-body invariant-mass, (left) hadron helicity angle, and (right) muon helicity angle, weighted by the inverse of the efficiency. The fitted angular decay rate including all its components is also shown.
    }
    \label{fig:fitprojections_massive}
\end{figure}

\begin{figure}[!ht]
    \centering
    \includegraphics[width=0.49\textwidth]{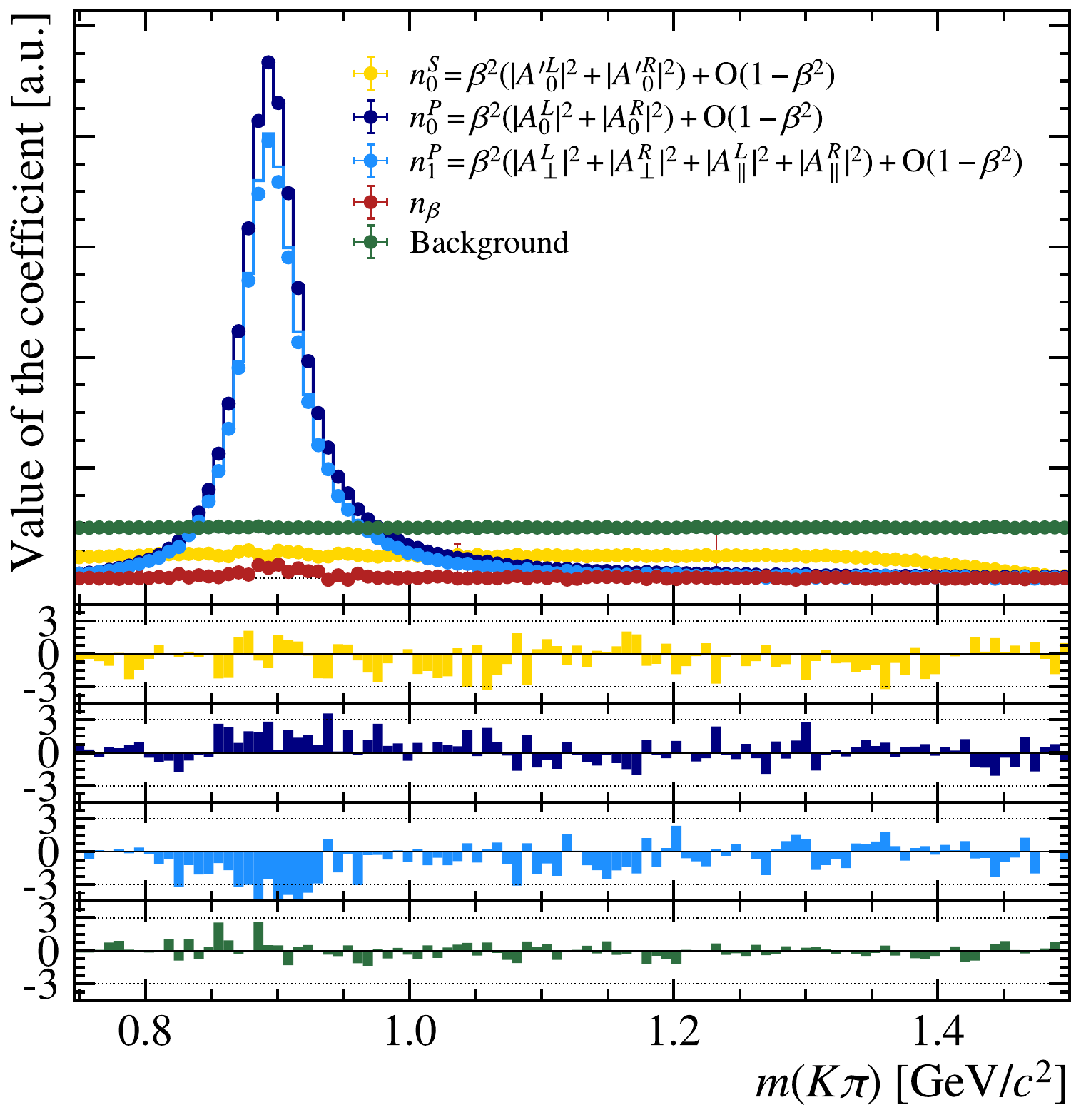}%
    \includegraphics[width=0.49\textwidth]{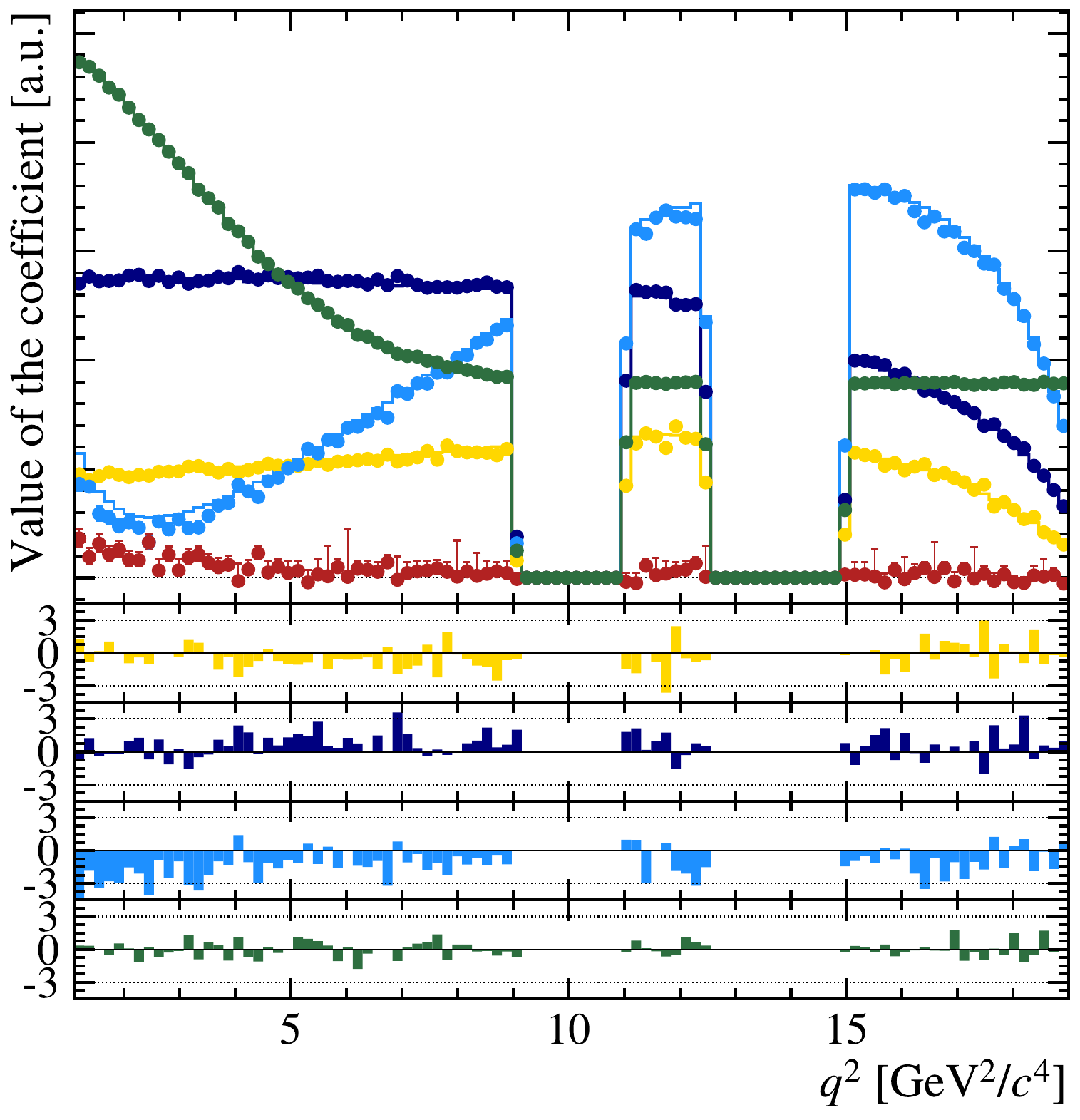}
    \caption{Extracted shapes (data points) compared to the generated shapes (lines) in the (left) dihadron invariant-mass and (right) dimuon invariant-mass squared for a large toy data set of more than 10 million data points, generated assuming massive muons.
    The faint dotted horizontal line indicates zero.
    The three bottom plots show the bin-wise pulls for the three components in their respective colour.
    No generated shape for $n_\beta$ is shown because the generation of a representative toy sample is non-trivial.
    Importantly, the small bias visible only in such a very large data set is not a measurement bias but can be entirely attributed to the difficulty of generating samples that accurately represent the $(n)'$ coefficients including the $b$-terms.}
    \label{fig:sweighted_highstats_massive}
\end{figure}
\clearpage
\printbibliography

\end{document}